\documentclass[11pt,reqno,a4paper]{amsart}
\usepackage[margin=2.5cm]{geometry}
\RequirePackage{amsthm,amsmath,amsfonts}
\RequirePackage[numbers]{natbib}
\RequirePackage[colorlinks,citecolor=blue,urlcolor=blue]{hyperref}
\usepackage{booktabs,longtable}
\usepackage{graphicx}
\usepackage{xcolor}
\usepackage{epstopdf}
\usepackage{physics}
\usepackage{bm}
\usepackage{hyperref}
\def\MR#1{\href{http://www.ams.org/mathscinet-getitem?mr=#1}{MR#1}}

\newcommand{\Expe}{{\rm \bf E}}
\newcommand{\Expp}{{\rm \bf P}}
\newcommand{\Var}{{\rm {\bf V}ar}}
\newcommand{\Cov}{{\rm {\bf C}ov}}

\newcommand{\bfx}{\boldsymbol{x}}
\newcommand{\bfX}{\boldsymbol{X}}
\newcommand{\given}{\, \vert \,}
\newcommand{\bftheta}{\boldsymbol{\theta}}
\newcommand{\bfmu}{\boldsymbol{\mu}}
\newcommand{\bfSigma}{\boldsymbol{\Sigma}}

\begin{document}
\begin{center}
	{\Large \bf  
	Functional data 
	analysis: An application to COVID-19 data in the United States}
	
	\bigskip
	{\bf Chen Tang\footnote{\scriptsize Research School of 
	Finance, Actuarial 
	Studies \& Statistics, The Australian National University, 
	Canberra, ACT 2601, AUS. Email: 
	\href{mailto:chen.tang@anu.edu.au}{chen.tang@anu.edu.au}} 
	\qquad Tiandong Wang\footnote{\scriptsize Department of 
	Statistics, Texas 
	A\&M University, College 
		Station, TX 77843, U.S.A. Email: 
		\href{mailto:twang@stat.tamu.edu}{twang@stat.tamu.edu}}
	\qquad Panpan Zhang\footnote{\scriptsize Department of 
	Biostatistics, Epidemiology and Informatics, University of 
	Pennsylvania, Philadelphia, PA 19104, U.S.A. Email: 
		\href{mailto:panpan.zhang@pennmedicine.upenn.edu}
		{panpan.zhang@pennmedicine.upenn.edu}
	\\ 
	${}^{\ast}${\bf All three authors equally contribute to this 
	research.}}
	}
		
	\bigskip
	
	\today
\end{center}

\bigskip\noindent
{\bf Abstract.}
The COVID-19 pandemic so far has caused huge negative impacts on 
different areas all over the world, and 
the United States (US) is one of the most affected countries. In 
this paper, we use  
methods from the functional data analysis to look into the 
COVID-19 data in the US. We explore the modes of 
variation of the data through a functional principal 
component analysis (FPCA), and study the canonical correlation 
between 
confirmed and 
death cases. In addition, we run a cluster analysis at 
the state level so as to investigate the relation between 
geographical locations and the clustering structure. Lastly, we 
consider a 
functional time series model fitted to the cumulative confirmed 
cases 
in the US, and make forecasts based on the dynamic FPCA. Both 
point and interval forecasts are 
provided, and the methods for assessing the accuracy of the 
forecasts are also included. 


Secondary: 62H30


\section{Introduction}
\label{sec:intro}

Ever since December 2019 when the outbreak of the Coronavirus 
Disease (COVID-19) has first 
come to the attention of the general public in 
China, this epidemic has been spread to more than 180 
countries over the world, leading to extremely negative impacts on 
global public health and economy. 
As of 08/25/2020, this virulent 
disease has caused a total of 
23,721,008 confirmed and 
815,029 death cases around the world, according to the data 
collected at 
Johns Hopkins University \cite{jhu:data}.
The United States (US) is the 
most severely hit country with 5,759,147 confirmed and 
177,873 death cases. According to the US Bureau of Labor 
Statistics, the unemployment rate of US has reached 
14.7\% in April 2020, a record-high over-the-month increase since 
January, 1948. Although the rate has declined to 8.4\% in August
2020, it still remains at a relatively high level, compared to less 
than 4\% before the pandemic. 

A great amount of scientific efforts have been integrated to learn 
the progression of the disease, and to mitigate its negative impact 
on people's normal life. However, according to some recent 
research~\cite{Grubaugh2020}, it is 
evident that the COVID-19 has become 
mutating. As pointed out by Dr.~Anthony Fauci, 
the director of the National Institute of Allergy and Infectious 
Diseases, on 06/24/2020, 
the vaccine for the COVID-19 would not likely to 
be available until 2021. On 07/27/2020, the National Institutes of 
Health has announced the phase 3 clinical trial of the 
investigational vaccine 
for the COVID-19. As currently there is no medication that 
can directly eliminate the virus, many infected patients 
have to rely on their own immune systems to 
recover under the help of standard treatments. As more 
and more COVID-19 data become available, statisticians now commit
to carrying out intensive data-driven analyses to precisely 
uncover the epidemic characteristics of the COVID-19.

In this paper, we exploit methods from the functional data analysis 
to analyze the COVID-19 data of both confirmed and death cases in the
US from 01/21/2020 to 08/15/2020. 
In this study, our data are collected at state level (see 
Section~\ref{sec:pre} for details), with the following research 
questions in mind:
\begin{itemize}
	\item[(1)] Does the practice of public health measures (e.g., 
	social 
	distancing and mask wearing) help to mitigate the spread 
	of the COVID-19? On the other hand, does the reopening of 
	business 
	exacerbate the spread of the disease?
	\item[(2)] Is there any quantitative way to understand the 
	correlation 
	between infections and deaths caused by the COVID-19 in the US? 
	Does the 
	correlation vary from state to state?
	\item[(3)] Is the spread of the COVID-19 related to the 
	geographical 
	locations of the infected regions or hot spots (at the state 
	level) in the US?
	\item[(4)] Is there a way to have some reasonable forecasts with 
	regard to the total number of 
	confirmed cases nationwide?
\end{itemize} 

\section{Preliminary analysis} 
\label{sec:pre}

We collect the number of confirmed and death cases of the COVID-19 
from the~50 continental states in the US between 01/21/2020 (the 
date of 
the first domestic confirmed 
case reported in the US) to 08/15/2020 (the weekend before school 
reopening in most of the states). The data of cumulative 
confirmed 
and death cases were collected at state 
level, from a publicly available repository released and updated 
by New York Times (\url{https://github.com/nytimes/covid-19-data}). 
In 
what follows, the daily confirmed and death cases are obtained 
effortlessly.

\begin{figure}[ht]
	\centering
	\includegraphics[width=\textwidth]{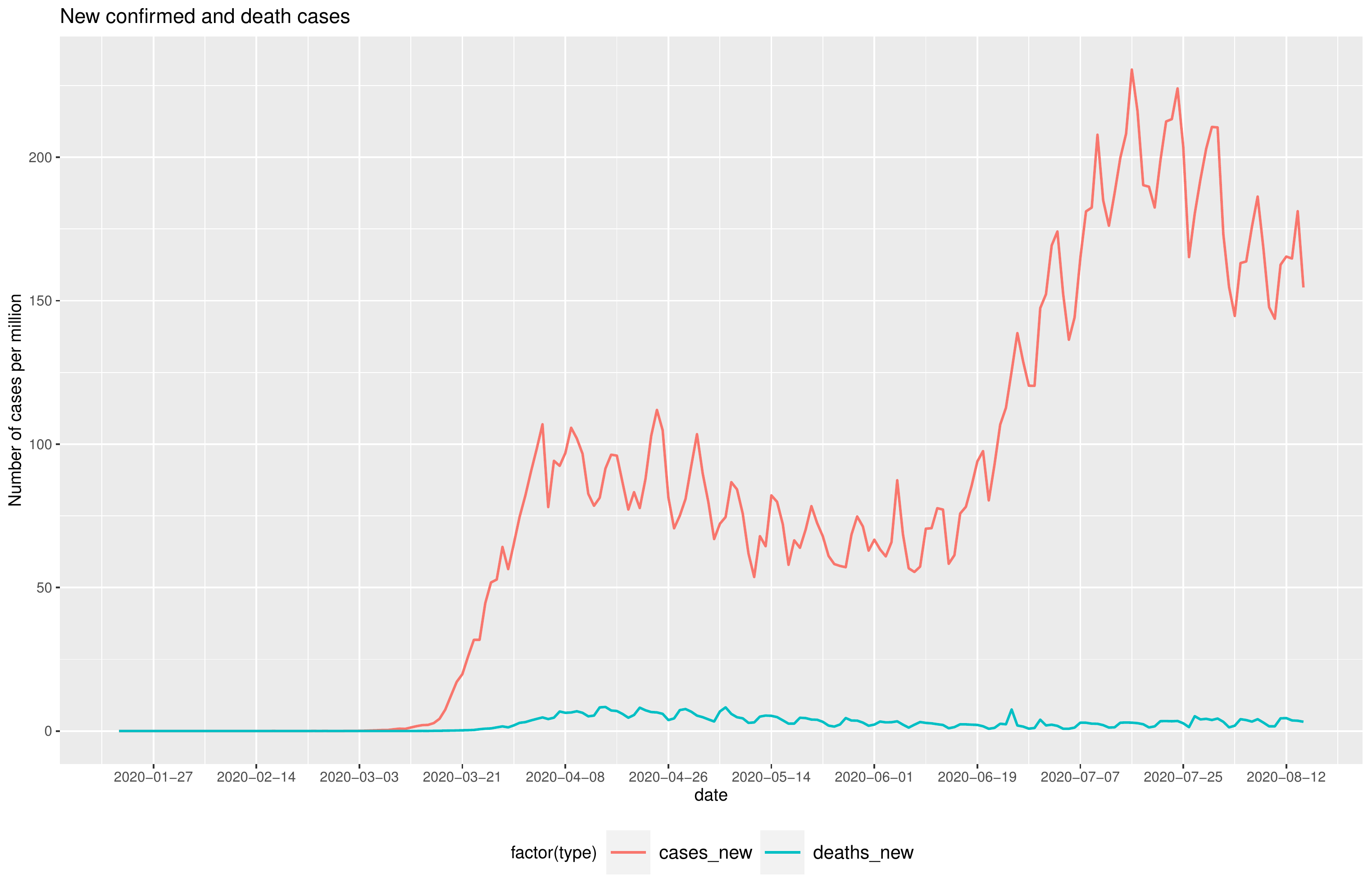}
	\caption{The number of daily confirmed and death cases (per 
		million) in the US from 01/21/2020 
		to 08/15/2020. }
	\label{Fig:usts}
\end{figure}

The majority of our analyses is done at the state level. Noticing 
the significant differences in
the population size for each state, 
we standardize the data, using the estimated population size for 
each state at 
the end of year 2019 from the US Census Bureau 
(\url{https://www.census.gov/popclock}) to scale all
collected data (cumulative and daily cases), and save them in units 
of ``per million''.
In Figure~\ref{Fig:usts}, we plot the number of daily confirmed 
and death cases in the US. From early May to early June, the 
cumulative 
case-to-fatality rate (CFR) of the COVID-19 in the US has stayed 
consistently 
high around 6.01\%, close to the estimate (6.1\%) given 
in~\cite{Abdollahi2020}, which is 
slightly higher than the CFR in Wuhan, China (5.8\%) reported on 
02/01/2020~\cite{Omer2020}, and significantly higher than the 
global 
CFR (3.4\%) according to WHO Director-General's opening remarks at 
the media 
briefing on 03/03/2020.  In August, the CFR in the US has declined 
to 3.1\%. 

One important epidemic metric for the study of the infectious 
disease dynamics is 
the basic reproduction number, usually denoted as $R_0$.
It refers to the average number of secondary cases per infectious 
case in 
a population where all the individuals thereof are susceptible to the
infection. A variant called the instantaneous reproduction 
number, denoted as $R_t$, is the average number of cases generated 
by each infection at a given time $t$. We observe that 33 
of 50 continental states in the US has had $R_t > 1$ 
(\url{https://rt.live}) on 05/29/2020, suggesting 
that the epidemic has not yet been fully contained by the end of 
May, 2020.  
In fact, this proportion climbs to 41 out of 50 in the following 
month. Upon the end of the study period, the proportion of the 
states with $R_t > 1$ has dropped to 16 out of 50, but the whole 
nation 
is still faced with the potential risk of a massive spread owing to 
the reopening of schools 
in the fall. Some other 
results of critical epidemic characteristics and 
dynamics of the COVID-19 in the US have been reported 
in~\cite{Peirlinck2020}.

\begin{figure}[htp]
	\centering
	\includegraphics[width=\textwidth]{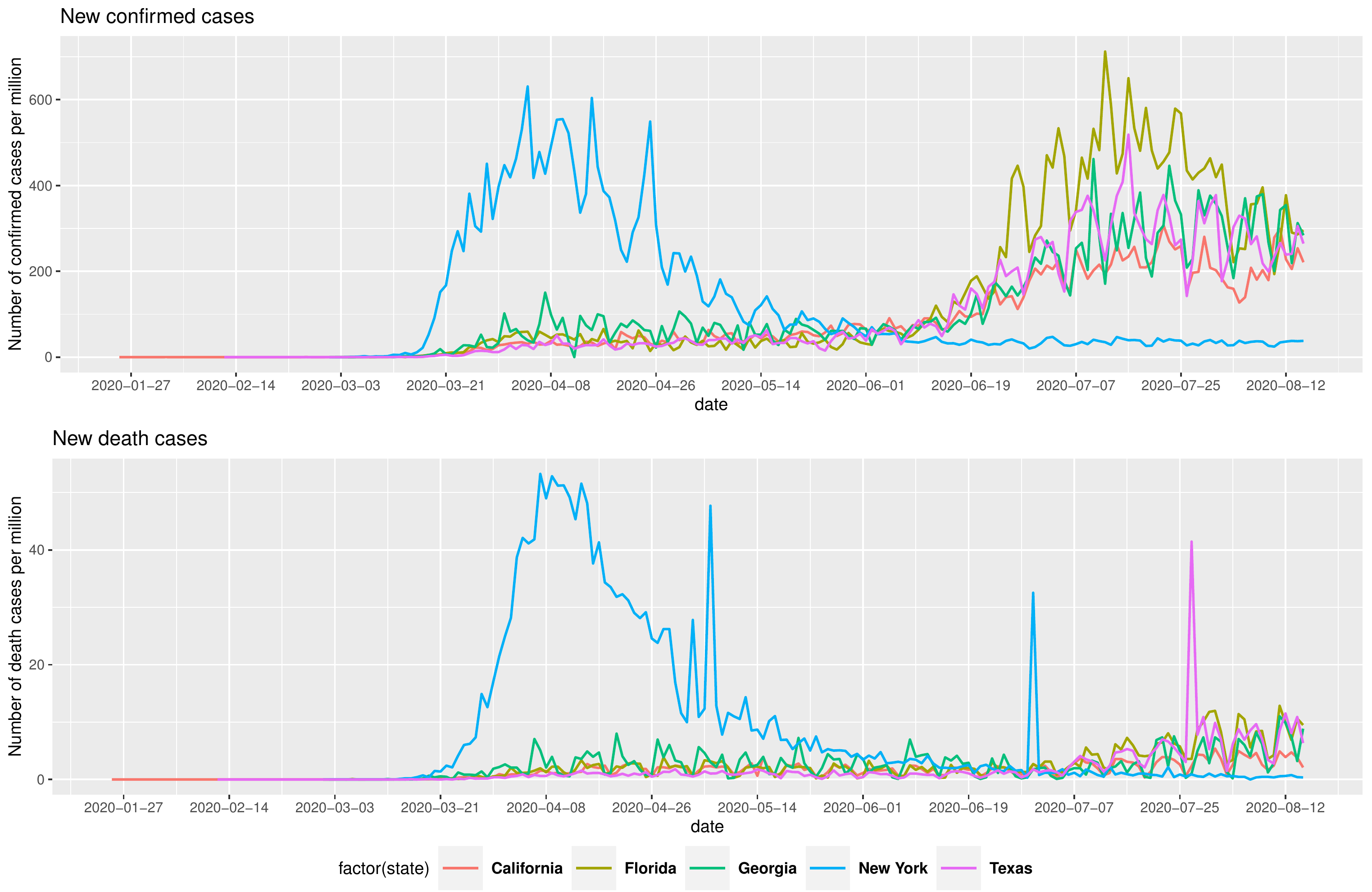}
	\caption{The number of daily confirmed and death cases (per 
		million) 
		of top five 
		continental states in the US.}
	\label{Fig:state}
\end{figure}

In this study, we treat the collected data as functional data, and 
adopt methods from the {\em 
	functional data analysis} (FDA).
Figure~\ref{Fig:state} shows the functional data of 
daily confirmed and death cases (per million) of the top five states 
in 
the
US during the study period. As discussed in~\cite{Ramsay1991}, the 
FDA methods are
applicable to sparsely and irregularly spaced data in
time, so are
preferred to the standard time series methods in the 
analyses of the time series data. Besides, the FDA methods manage to
capture the functional behavior of the underlying data which 
generate the
process (see~\cite{Ramsay1982,Ramsay1991,Ramsay2002} for details), 
and have been widely adopted in a plethora of applications 
in public 
health and biomedical studies~\cite{Ullah2013}.
In the next few
sections, we list the adopted FDA methods, and present the analysis 
results when they are applied to the COVID-19 data in the US.

\section{Functional principal component analysis and modes of 
	variation}\label{sec:fpca}

In multivariate statistics, {\em modes of variation} are a set of
centered vectors describing the variation in a population or sample. 
Typically, 
variation patterns are characterized via the standard eigenanalysis, 
i.e., principal component analysis~\cite{Jolliffe2002}. Analogously 
in the FDA, modes of variation provide an 
efficient tool to visualize the variation of the functional curves 
around the mean function. Identifying modes of variation in 
functional data is usually done through the functional principal 
component analysis~\cite{Dauxois1982}, providing new 
insights and precise 
interpretations of the functional data.

\subsection{Functional principal component analysis}
\label{subsec:fpca}

Consider a 
probability space $(\mathbb{R}_+, \mathcal{F}, \Expp)$, and 
a compact interval $I\subset\mathbb{R}_+$. A stochastic 
process $X(\cdot)$ is called an $L_2$ process on~$I$ if
\begin{equation*}
	\Expe \int_I |X(t)|^2 \, dt < \infty.
\end{equation*}
Let $X_i$, for $i = 1, 2, \ldots, n$, be
independent realizations of the underlying $L_2$ process 
$X(\cdot)$.
By convention in the FDA, functional 
data are given as
\begin{equation} \label{eq:fdata}
	Y_{i}(t) = X_{i}(t) + \varepsilon_{i}(t),
\end{equation}
for $t \in \mathcal{T}_i := \{t_{ij}, j = 1, 2, \ldots, n_i\}$, a 
time schedule for 
subject $i$. The terms $\varepsilon_{i}(t_{ij})$'s are independent 
measurement 
errors with 
$\Expe(\varepsilon_{i}(t_{ij})) = 0$ and 
$\Var(\varepsilon_{i}(t_{ij})) = 
\sigma_{ij}^2$ for some constant $\sigma_{ij}$.

The {\em functional principal component analysis} (FPCA) is a
powerful tool for dimension reduction in the 
FDA~\cite{Dauxois1982}. In 
essence, FPCA is an expansion of the realization $X_i(t)$ into 
functional bases consisting of the eigenfunctions of the 
variance-covariance structure of the process $X(\cdot)$, where the 
eigenfunctions are required to be {\em orthogonal}. Let $\{\phi_k : 
k \ge 1\}$ denotes the collection of orthogonal eigenfunctions. In 
addition,
let $\mu(t) = \Expe(X(t))$ be the true mean function, and $\xi_{ik} 
:= 
\int_I (X_i(t) - \mu(t)) \phi_k(t) \, d t$ be the $k$-th {\em 
	functional principal component} (FPC) of $X_i$ 
(also called {\em scores} in the jargon).
By the {\em Karhunen–Lo\`{e}ve 
	Theorem}~\cite{Karhunen1946spektraltheorie, 
	Loeve1955probability}, we have
\begin{equation}
	X_{i}(t) = \mu(t) + \sum_{k = 1}^{\infty} \xi_{ik} 
	\phi_{k}(t).
	\label{eq:fpca}
\end{equation}
Due to the difficulties in estimating and interpreting the infinite
sum in Equation~\eqref{eq:fpca}, a conventional treatment is to 
approximate it by a finite sum of $K$ terms. In what follows, we set
\begin{equation*}
	Y_{i} (t)= \mu(t) + \sum_{k = 1}^{K} \xi_{ik} \phi_k(t) + 
	\upsilon_{i}(t),
\end{equation*}
where $\upsilon_{i}(t)$ is the counterpart of $\varepsilon_{i}(t)$ 
in Equation~\eqref{eq:fdata} 
owing to truncation.
We 
refer interested readers to~\cite{Shang2014} for a comprehensive 
review of the 
fundamental theory of FPCA. One primary 
application of FPCA is to explore modes of 
variation~\cite{Jones1992} for the
functional data, reflecting the percentage of total variations 
contributed by each principal eigenfunctions. 

When applying the FPCA method to the COVID-19 data of the 50 
continental states in the US, we notice that they are not identical 
in 
time schedules. The identification of the first COVID-19 case may 
differ by as long as 
30 days across the country. Having observed the sparsity in the
functional data during the early stage of the outbreak, 
we adopt the {\em Principal Components 
	Analysis through Conditional Expectation} (PACE) approach 
proposed 
in~\cite{Yao2005} for our analysis. The estimation of the FPC score 
of $\bm{Y}_i = (Y_{i}(t_{i1}), \ldots, 
Y_{i}(t_{in_{i}}))^{\sf T}$, for $i = 1, 2, \ldots n$, via PACE 
takes
the following major steps:
\begin{itemize}
	\item[(1)] For each $\bm{Y}_i$, estimate the 
	mean function 
	$\bm{\mu}_i = (\mu(t_{i1}), 
	\ldots,\mu(t_{in_{i}}))^{\sf T}$ and the covariance structure 
	$\bm{\Sigma}_{\bm{Y}_i} = \Cov(\bm{Y}_i, 
	\bm{Y}_i)$ by locally
	linear scatter and surface
	smoothers, respectively~\cite{Fan1996};
	\item[(2)] Discretize the off-diagonal 
	smoothed covariance to estimate the eigenfunctions 
	$\bm{\phi}_{ik} = 
	(\phi_k(t_{i1}), \ldots, \phi_k(t_{in_{i}}))^{\sf Tc}$, $k = 1, 
	2, 
	\ldots 
	K$,  and the 
	corresponding
	eigenvalue $\lambda_k$~\cite{Capra1991};
	\item[(3)] Adopt an Akaike Information Criterion (AIC)
	type criterion to 
	select the 
	number of eigenfunctions, i.e., $K$, needed to approximate the 
	process~\cite{Shibata1981};
	\item[(4)] Lastly, the estimation of the FPC score for the 
	$i$-th 
	subject is given through the 
	conditional expectation:
	\begin{equation}\label{eq:xi}
		\hat{\xi}_{ik} = \Expe(\xi_{ik} \, | \, \bm{Y}_i) = 
		\hat{\lambda}_k 
		\hat{\bm{\phi}}^{\sf 
			T}_{ik}\hat{\bm{\Sigma}}^{-1}_{\bm{Y}_i}(\bm{Y}_i - 
		\hat{\bm{\mu}}_i),
	\end{equation}		
	where $\hat{\lambda}_k$, $\hat{\bm{\phi}}_{ik}$, 
	$\hat{\bm{\Sigma}}_{\bm{Y}_i}$ and $\hat{\bm{\mu}}_i$ 
	are respectively the estimates of $\lambda_k$, 
	$\bm{\phi}_{ik}$, $\bm{\Sigma}_{\bm{Y}_i}$ and 
	$\bm{\mu}_i$.
\end{itemize}

In practice, the {\tt R} package 
{\bf fdapace}~\cite{Carroll2020} allows us 
to apply the PACE method to the functional COVID-19 data directly.
The computation results and corresponding discussions are given in 
the subsequent section.

\subsection{Modes of variation}
\label{sec:modes}

\begin{figure}[tbp]
	\centering
	\includegraphics[width=\textwidth]{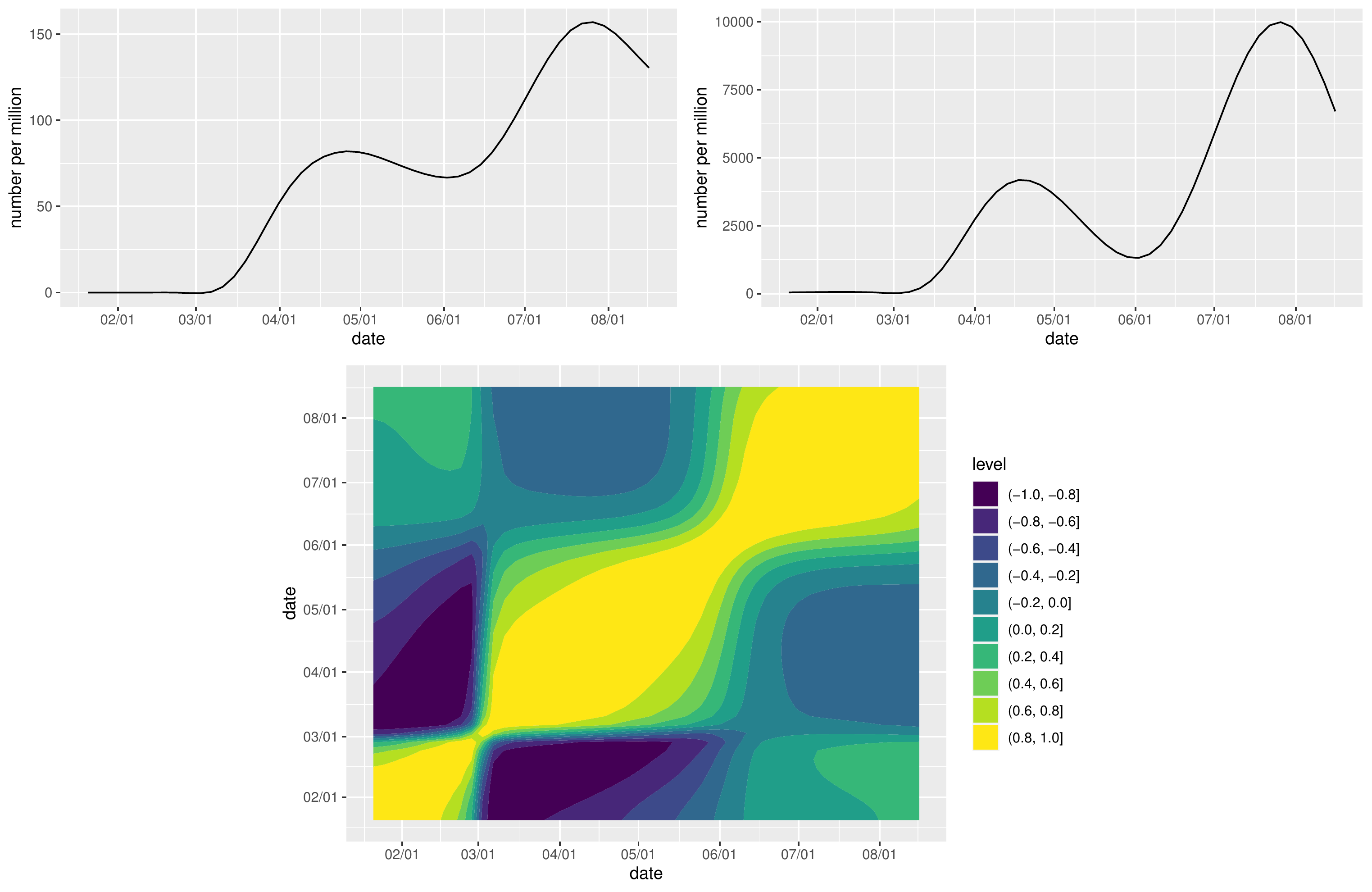}
	\caption{Fitted mean curve (top left panel), fitted variance 
		curve (top right panel) and fitted 
		correlation surface (bottom) of daily confirmed cases of 
		COVID-19 in the US.}
	\label{Fig:fitted}
\end{figure}

To begin with,
we plot the fitted mean curve (which 
estimates the trend over time), the fitted variance curve  (which 
estimates the subject-specific variation)
and the fitted covariance surface of daily confirmed cases across 50 
continental states in Figure~\ref{Fig:fitted}.

The estimated mean 
is close to $0$ in January and February, and starts climbing since 
early March until reaches a local maximum in late April. The overall 
trend in May is downwards, followed by a second wave that starts
from June. The curve hits the peak at the end of the third week 
of July. The estimated variance curve looks similar to the fitted 
mean curve in shape, which stays low at the very early 
stage, 
and deviates from $0$ at the end of the first week of March. The 
first local maximum emerges around 04/17/2020, and then starts 
decreasing until 06/01/2020. The global maximum of the curve is 
observed in the third week of July, corresponding to the worst 
period of the attack of the COVID-19 to the country. The correlation 
surface is 
presented through a contour plot. Measurements at close time 
points appears to be highly correlated. The correlation between 
early and very late times are close to $0$, whereas that
between early and middle times tends to be negative. The correlation 
patterns among middle and later times are slightly negative. In 
particular, the correlation surface reveals that
the increase in the number of confirmed cases follows three 
different stages, where the two break points are respectively 03/01
and 06/01. The strong positive correlation 
seems more persistent in the latter two stages.

\begin{figure}[ht]
	\centering
	\includegraphics[width=\textwidth]{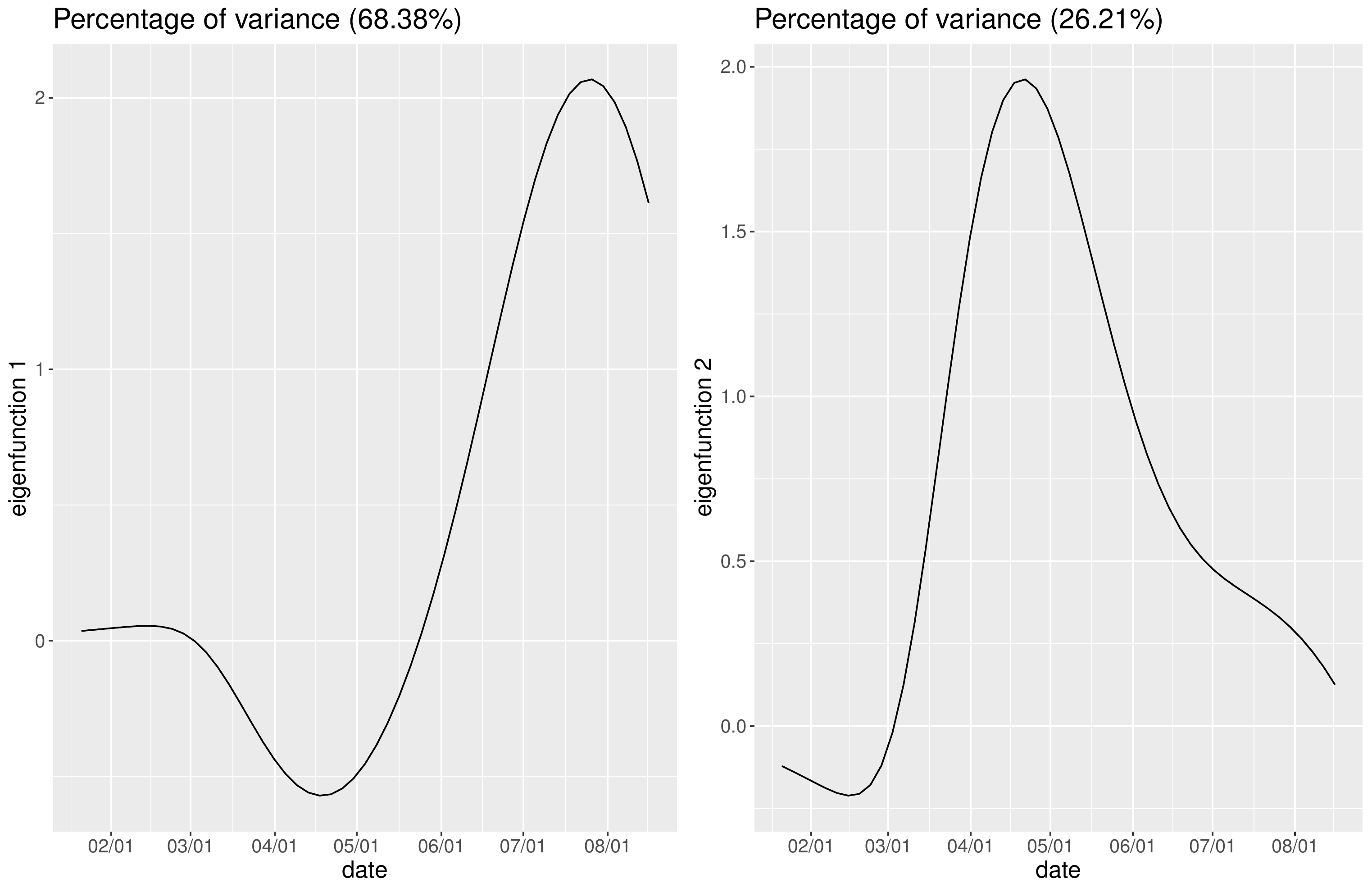}
	\caption{First two eigenfunctions for the US COVID-19 confirmed 
		cases}
	\label{Fig:eigenfun}
\end{figure}

Next we apply the PACE method to the COVID-19 data of daily 
confirmed cases in the US. The adaptive algorithm selects a total of 
six eigenfunctions (accounting for more than 99.99\% of the total 
variation), where the first two together explains about 95\% of the 
total variation of the data (see Figure~\ref{Fig:eigenfun}), 
indicating
the remaining eigenfunctions are less important.

The first eigenfunction (accounting for 68.38\% of the total 
variation) remains constant close to $0$ during the first month of 
the 
study period, and starts decreasing afterwards. The value of the 
eigenfunction 
falls negative since 03/01/2020 until it reaches a local minimum 
around 04/17/2020. After that, the function value keeps increasing 
until it hits the peak around 07/23/2020. The first eigenfunction 
defines the most important mode of variation, suggesting that the 
state-specific intercept term captures the vertical shift 
(especially later times) in the 
overall mean. Besides, the first eigenfunction reflects a 
contrast between middle 
and late times, where the break point emerges about one week after 
the announcement of business reopening in the majority of the states 
in the US.

The second eigenfunction presents a variation around a piecewise 
linear time trend. The value of the eigenfunction is negative at 
early times, and 
becomes positive since March, and hits a break point 
in the third week of April. Since then, it begins to decrease but 
stays positive during the rest of the study 
period. The second eigenfunction suggests that the second largest 
variation 
among the states is a scale difference along the direction of this 
functional curve.

\section{Functional canonical correlations}
\label{sec:fcca}

In statistics, canonical correlation analysis is a 
tool to make
inference based on the cross-covariance matrix of
multiple datasets. The analogue in FDA, named {\em functional 
	canonical correlation 
	analysis} (FCCA), aims to investigate the correlation 
shared by multiple functional datasets~\cite{Leurgans1993}. 
Specifically, here we use the FCCA method to quantitatively estimate 
the correlation between confirmed and death cases across the states. 

\subsection{Functional canonical correlation analysis}
\label{subsec:fcca}

We first outline the theoretical setup of the analysis 
method~\cite{He2003, He2004}, and refer the interested readers 
to~\cite{Wang2016functional} for a concise review.
Let $X(\cdot)$ and $Y(\cdot)$ be two $L_2$ 
processes on 
two compact intervals, $I_X$ and $I_Y$, respectively. In 
addition, let $\mathcal{H}_2(I)$ denote the 
{\em Hilbert space} of square-integrable functions on some compact 
interval $I$,
with respect to Lebesgue measure. In addition, the notation 
$\langle \cdot, 
\cdot \rangle$ refers to the standard operation of inner product. By 
definition, the first 
canonical 
correlation is given by
\begin{equation*}
	\rho_1 = \sup_{u \in \mathcal{H}_2(I_X), v \in 
	\mathcal{H}_2(I_Y)}
	\Cov\bigl(\langle u, X \rangle, \langle v, Y \rangle\bigr) 
	= \Cov\bigl(\langle u_1 , X \rangle, \langle v_1 , Y 
	\rangle\bigr),
\end{equation*} 
subject to
\begin{equation}
	\label{eq:fccacond}
	\Var\bigl(\langle u , X \rangle\bigr) = 1 \qquad \mbox{and} 
	\qquad \Var\bigl(\langle v , Y \rangle\bigr) = 1. 
\end{equation}
The pair of the optimal solutions $(u_1, v_1)$ are called the {\em 
	canonical weight functions} of $\rho_1$. For $k \ge 2$, the 
$k$-th 
canonical 
correlation coefficient, denoted 
$\rho_k$, 
is defined under the condition of orthogonality:
\begin{equation*}
	\Cov\bigl(\langle u_k , X \rangle, \langle u_j , X 
	\rangle\bigr) = \Cov\bigl(\langle v_k , Y \rangle, \langle 
	v_j , Y \rangle\bigr) = 0,\qquad j = 1, 2, \ldots, k - 1.
\end{equation*}
Set
$\rho_k$, $k\ge 2$, and its associated weight functions $(u_k, v_k)$ 
in an analogous way:
\begin{equation*}
	\rho_k = \sup_{u \in \mathcal{H}_2(I_X), v \in 
	\mathcal{H}_2(I_Y)}
	\Cov\bigl(\langle u, X \rangle, \langle v, Y \rangle\bigr) 
	= \Cov\bigl(\langle u_k , X \rangle, \langle v_k , Y 
	\rangle\bigr),
\end{equation*}
subject to condition~\eqref{eq:fccacond}. In FDA, the  inner 
products 
$\langle u, X\rangle$ and $\langle v, 
Y\rangle$ corresponding to weight functions $u$ and $v$ are called 
{\em probe scores}. Typically for $k \ge 1$, we refer to 
$(u_k, v_k)$ as the canonical weight 
pair that optimizes the canonical criteria for $\rho_k$.

The FCCA 
approach, in essence, determines the projection of $X$ in the 
direction of 
$u_k$ as well as that of $Y$ in the direction of $v_k$, such that 
their linear combinations are maximized for $k \ge 1$. Consider 
three kinds of variance and cross-covariance operators $\Sigma_{XX} 
: \mathcal{H}_2(X) \mapsto \mathcal{H}_2(X)$,  $\Sigma_{YY} : 
\mathcal{H}_2(Y) 
\mapsto \mathcal{H}_2(Y)$ and $\Sigma_{YX} : \mathcal{H}_2(Y) 
\mapsto 
\mathcal{H}_2(X)$, which correspond to the variance structure 
of $X$, the variance structure of $Y$ and the cross-covariance 
structure between $X$ and $Y$, respectively. 
We then rewrite the definition expression of $\rho_1$ as follows:
\begin{equation}
	\label{eq:rewriterho}
	\rho_1 = \sup_{u \in \mathcal{H}_2(I_X), v \in 
	\mathcal{H}_2(I_Y)} 
	\langle 
	u, \Sigma_{YX}\, v \rangle,
\end{equation}
subject to $u^\top \Sigma_{XX} u = v^{\top} \Sigma_{YY} v = 1$. 
Similar arguments can also be applied to~$\rho_k, k \ge 2$, by 
accounting for the condition of orthogonality. A standard procedure 
for the change of 
basis in Equation~(\ref{eq:rewriterho}) implies that the problem is 
equivalent to an eigenanalysis of $\Sigma_{XX}^{-1/2} \Sigma_{YX} 
\Sigma_{YY}^{-1/2}$, i.e., a maximization problem of {\em Rayleigh 
	quotient}.

In this study, we use functions 
from the {\tt R} package {\bf fda}~\cite{Ramsay2020} to implement 
the FCCA method based on an integration of an exceedingly greedy 
procedure and an expansion of the functional basis. 
A variety of methods solving the optimization problem of 
functional canonical correlations have been developed in the 
literature; see for example,~\cite{Ramsay2002, He2003, 
	He2004, Yang2011}.

Prior to estimating 
the functional canonical correlation between confirmed cases
and death tolls in the US, some additional pre-processing procedures 
to 
the data are necessary, as we observe that the date on which the 
first confirmed case is reported varies significantly
across the states, and the number of death counts stays
relatively low during the entire study period in several states. 

\subsection{Canonical correlations between confirmed and 
	death cases}
\label{subsec:fccaus}

Now we inspect the functional canonical correlation 
between confirmed and death cases from the 50 continental 
states in the US. The first step is to 
pre-process the data. Noticing that diagnostic tests (e.g., 
swab test) of the
COVID-19 have not been widely implemented in most states until 
late March (for instance, the first drive-through COVID-19 testing 
site in Philadelphia was not open to the public until 03/20/2020), 
we reschedule the starting date of study period to 04/01/2020 for 
the rest of the study in this section. 
Besides, the cumulative number of (scaled) confirmed and death cases 
(instead of scaled daily numbers) are used for the analysis so that 
no state is 
excluded due to sparsity, especially for those with few cases 
reported 
during April and early May.

\begin{figure}[ht]
	\centering
	\includegraphics[width=\textwidth]{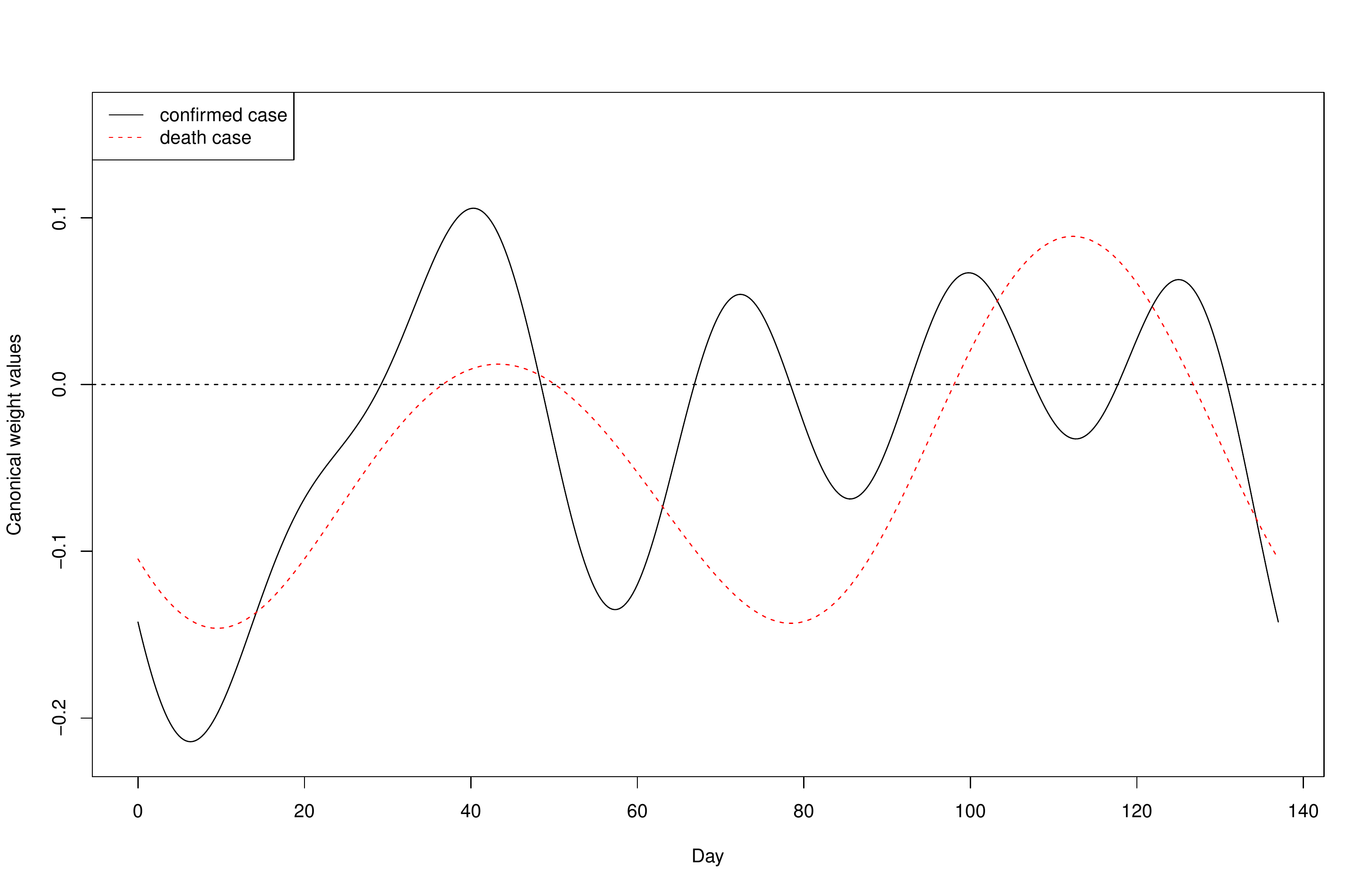}
	\caption{The first pair of canonical weight functions 
		correlating confirmed and death cases of the 50 
		continental 
		states in the US. }
	\label{Fig:cwt}
\end{figure}

In the literature, there are several different options of basis 
functions 
for the basis expansion in the FCCA. Having observed some periodic 
features in our functional data, we choose the Fourier basis here. 
The 
first canonical correlation is $0.985$, which reveals a dominant 
pair 
of modes of variation that are highly positively correlated. The 
corresponding 
canonical weight functions are plotted in Figure~\ref{Fig:cwt}. The 
canonical weight function of confirmed cases resembles a sinusoidal 
function with a period of one month approximately, while the 
counterpart of 
death cases seemingly contrasts early and middle times to late 
times (primarily in July). A state has a negative score with a large 
absolute value
on the canonical 
weight of confirmed cases if it has a large number of confirmed 
cases in 
April, late May and early June, but 
small counts in July and early August. Also, a state has 
a high positive score on 
the canonical weight of death cases if it has more death cases in 
July than any other time during the study period.

\begin{figure}[ht]
	\centering
	\includegraphics[width=\textwidth]{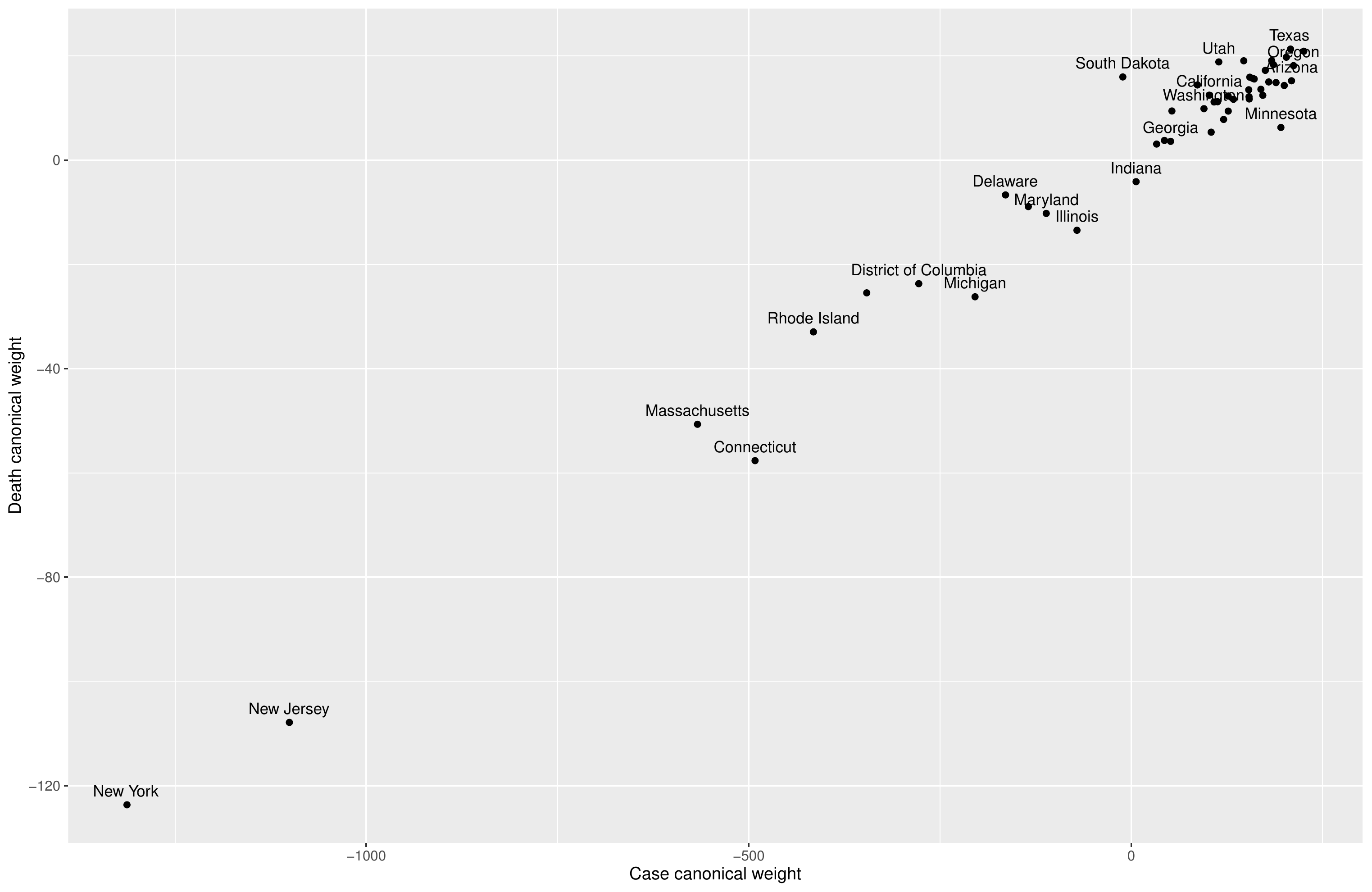}
	\caption{Canonical variable scores of death 
		cases against confirmed cases of the 50 continental states 
		in 
		the US.}
	\label{Fig:fcca}
\end{figure}

Next we plot the canonical variable scores for the death cases 
against confirmed cases in Figure~\ref{Fig:fcca}, 
where we see that the New York state lies on the 
bottom left corner. This is due to the fact that the New York state 
(especially New York City) has a lot of confirmed and death cases at 
the early stage of the pandemic, but 
later the 
transmission of the virus has been well controlled since July, 
thanks 
to the complete shutdown of the city as well as the rigorous 
adherence 
to public health measures, e.g. the practice of social distancing, 
mask wearing and the limit of the number of people allowed in 
essential businesses. Since July, less than 20 deaths has been 
reported daily in 
New York City, in contrast with an average of more than 700 deaths 
every day in 
April. Therefore, both canonical variable scores remain low in the 
state 
of New York. 

Another ``outlier'' is New 
Jersey, which is geographically adjacent to New York state. The 
overall 
trends of confirmed and death cases in New Jersey are similar in 
shape to those in New York, but both are 
smaller in magnitude. Therefore, New Jersey sits close to New York 
in 
Figure~\ref{Fig:fcca}. Three states in the middle with negative 
scores 
in both canonical variables are Massachusetts, Connecticut and Rhode 
Island, all of which are geographically close to New York. 

All other states are mostly scattered on the top right corner in 
Figure~\ref{Fig:fcca}. The one sits at the
most top right is Texas, with positive scores in both canonical 
variables. 
The confirmed and death cases in Texas are not the 
largest in the first wave of outbreak (i.e., April and May), but the 
state
has seen a large surge in the daily confirmed and death 
cases since late June. Consequently, both of the canonical scores of 
Texas remain positive. 
In fact, Texas has controlled the spread of the disease well in 
April 
when a 14-day self-quarantine has been mandated, 
all nonessential 
businesses are closed,
and ordered travel 
restrictions have been instituted between Texas and Louisiana (an 
outbreak 
occurs in New Orleans during that period). 
However, with the stay-at-home order lapsed in 
May, and the reopening of the economy started in June, Texas has 
experienced a
huge increase in the COVID-19 cases, due to the
increasing frequencies of indoor and outdoor gatherings in large 
groups of people
without proper public health measures implemented.

\section{Functional cluster analysis}
\label{sec:clust}

Clustering is another common tool for data exploration in 
multivariate statistics, aiming at
constructing homogeneous groups (called clusters) 
consisting of observed data that present some similar 
characteristics or patterns~\cite{Kaufman1990}. In contrast,  
observations from different groups are expected to be as 
dissimilar as possible. The {\em functional cluster analysis} (FCA) 
is an unsupervised learning process for 
functional data. In this section, we investigate the cluster 
structure in the US based on cumulative confirmed cases (after being 
scaled) across the states.

\subsection{Methods for functional cluster analysis}
\label{subsec:fca}

Two classical methods for the FCA are hierarchical 
clustering~\cite{Ferreira2009} and 
$k$-means clustering~\cite{Abraham2003}. More recently, many 
clustering methods extended from them were proposed. 
See~\cite{Jacques2013} for a summary.

Traditional $k$-means clustering methods for the FCA, 
e.g.~\cite{Abraham2003}, require a finite set of 
pre-specified basis functions in order to span a functional space, 
and assume the observed functional data to admit the basis 
expansion. The FCA is then completed by applying the standard 
$k$-means algorithm to the estimated basis coefficients. 
Alternatively, one may replace basis coefficients with FPC 
scores (i.e., $\xi_{ik}$'s in Equation~\eqref{eq:xi}) for conducting 
the cluster analysis~\cite{Chiou2007, Chiou2008, Peng2008}.

In modern FDA, there are two methods that are extensively popular 
to conduct the FCA. One approach is the EM-based 
algorithm~\cite{Chen2015}, which
assumes a 
finite (say $K \in  \mathbb{N}$) mixture of Gaussian distributions. 
The model is given by
\begin{equation}
	\label{eq:mixgauss}
	f(\bfx \given \bftheta) = \sum_{c = 1}^{K} \pi_c \psi(\bfx 
	\given 
	\bfmu_c, \bfSigma_c),
\end{equation}
where $\bftheta = \{\pi_1, \ldots, \pi_K, \bfmu_1, \ldots, 
\bfmu_K, \bfSigma_1, \ldots, \bfSigma_K\}$ is a collection of 
parameters; for each $c = 1, 2, \ldots, K$, $\pi_c \in (0, 1)$ is 
the 
mixing proportion 
such that $\sum_{c = 1}^{K} \pi_c = 1$, and $\psi(\bfx \given 
\bfmu_c, \bfSigma_c)$ is a Gaussian distribution with mean $\bfmu_c$ 
and variance $\bfSigma_c$. The {\em log-likelihood} of the model in 
Equation~\eqref{eq:mixgauss} is optimized by the EM 
algorithm~\cite{Dempster1977}. The setups of the E-step and the 
M-step are standard, available in a variety of articles, 
texts and tutorials, e.g.~\cite{Lee2012}. Given 
the 
number of clusters $K$, 
the EM algorithm partitions a set of $n$ observations $\bfX = 
\{\bfx_1, 
\ldots, \bfx_n\}$ into $K$ clusters by maximizing the following 
expression:
$$\underset{c}{\rm arg \, max} \; \hat{\pi}_c \phi(\bfx_i \given 
\hat{\bfmu}_c, \hat{\bfSigma}_c)$$
for each $i = 1, 2, \ldots, n$. 

The alternative is 
the {\em $k$-centers functional clustering} (kCFC) algorithm 
proposed by~\cite{Chiou2007}, using 
the subspace spanned by the FPCs as cluster 
centers. 
This approach is popular as the distributional assumptions are 
relaxed.

Nonetheless, there are limitations in both algorithms. The kCFC 
algorithm assumes 
equal within-cluster variance, whereas the EM algorithm assumes a 
mixture of Gaussian distributions. Relevant discussions on
the difference between the two algorithms are included
in~\cite{Jacques2013}. In the present 
analysis, it seems inappropriate to assume equal within-cluster 
variance, so we adopt the model-based EM algorithm, which is 
available 
in {\tt R} package 
{\bf EMCluster}~\cite{Chen2015}. Since the method is based on the 
EM 
algorithm, it is critical to select the initial values of the 
parameters appropriately. Specifically, we adopt a strategy 
developed in~\cite{Biernacki2003} for the initialization of the 
algorithm, with corresponding functions available in 
{\bf EMCluster}. 

\subsection{Cluster analysis results}
\label{sec:clustres}

\begin{figure}[ht]
	\centering
	\includegraphics[width=\textwidth]{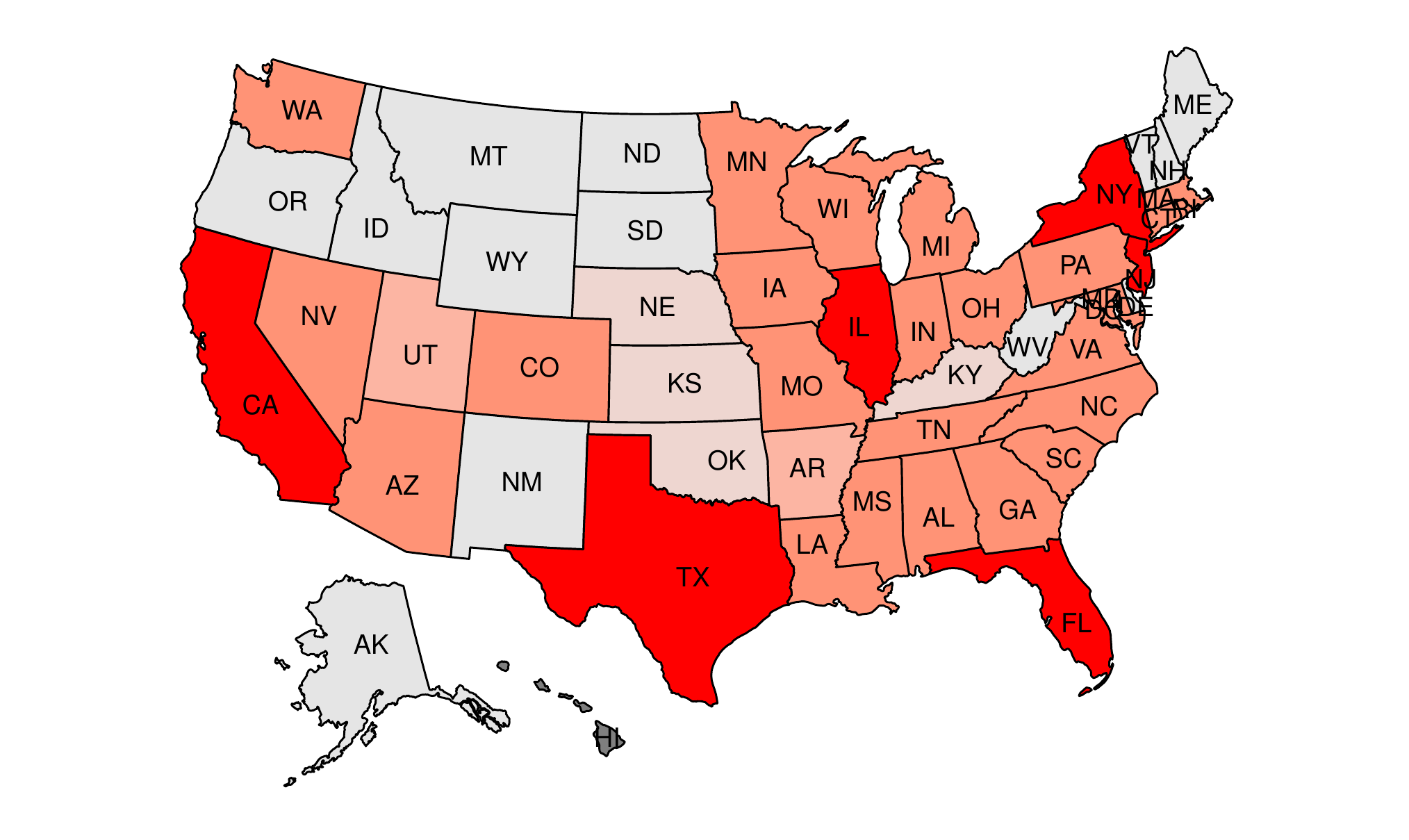}
	\caption{Clustering results based on the number of cumulative 
		confirmed cases in 50 continental states in the US from 
		04/01/2020 to 08/15/2020.}
	\label{Fig:cluster}
\end{figure}

Similar to the FCCA part, we trim the study period to ``04/01/2020 
to 08/15/2020'' as few (or even no) confirmed case has been observed 
in many states before April. The implementation of the EMCluster 
algorithm requires the predetermination of the number 
of clusters in prior. There 
are a couple of standard methods to obtain an optimal value of 
$K$. We select the value of $K$ via the {\em elbow method} based 
on the sum of squares of the within-cluster variations. The 
computation 
reveals that the most appropriate value of $K$ is $5$, and the 
optimal 
clustering strategy associated with $K = 5$ is given in 
Figure~\ref{Fig:cluster}.

The first cluster includes New York, New Jersey and Illinois, the 
three states that have been most severely attacked by the COVID-19 
in the 
first wave, as well as Florida, Texas and California, all of which 
have experienced a huge surge in both confirmed and death cases in 
the second wave. Overall, 
these six states are the worst hit by the COVID-19 throughout the 
study period.

The second cluster contains 24 states, including Georgia, North 
Carolina, Massachusetts, Pennsylvania and Washington, etc., most of 
which are geographically close (but not 
necessarily 
adjacent) to the hot-spot regions. Some states, e.g. Georgia, North 
Carolina and Colorado, actually have reached the peak of the
COVID-19 cases in the second wave after the reopening of the 
business,
when some mandatory protocols have been called off. Besides, with 
the 
relaxation of travel restrictions, an increase in population 
movements may also cause the cross infections among the residents 
in these states. The state of Washington, located at the northwest 
corner, 
also belongs to this cluster, where the first case in the US has 
been reported.

The third cluster only has two states: Utah and Arkansas. Utah is 
not directly adjacent those epidemic centers specified in the first 
cluster. The control of the
COVID-19 remains reasonably well in Utah, and we speculate some 
possible reasons as follows.
Though there is no evidence that youngsters are not likely 
to be infected, they are likely to have stronger immune systems 
(than elders). Utah has the lowest percentage of residents above 65 
in 
the nation. Besides, Utah is among the healthiest states in the 
nation overall 
(\url{https://www.americashealthrankings.org}). The attack of the 
COVID-19 
to the other state in the third cluster, Arkansas, 
has also been less serious during the second wave, although it is 
right 
next to Texas. The local government has maintained a series of 
policies 
that help reduce the spread of the disease, for example, the delay 
of school opening. 

The fourth cluster contains Oklahoma, Kansas, Nebraska and Kentucky, 
where 
the major industries are agriculture, core mining, 
and aviation, etc. Besides, there are not too many tourist resorts 
in these states triggering large population movements. It is also 
necessary to note that the population densities are relatively small 
in these states. Hence, they are less affected by the COVID-19 
compared to those in the first three clusters. 

Lastly, a total of $14$ states are classified in the fifth cluster, 
including Vermont, New Mexico, Alaska, Montana, and Wyoming, etc. 
These states are the least attacked by the virus. Although the data 
has been scaled, most of these states are large in 
their
geographical sizes, leading to low population densities. In fact, 
among 
all 50 states, 
Alaska, Montana and Wyoming have the three lowest 
population densities in the nation.

\begin{figure}[ht]
	\centering
	\includegraphics[width=\textwidth]{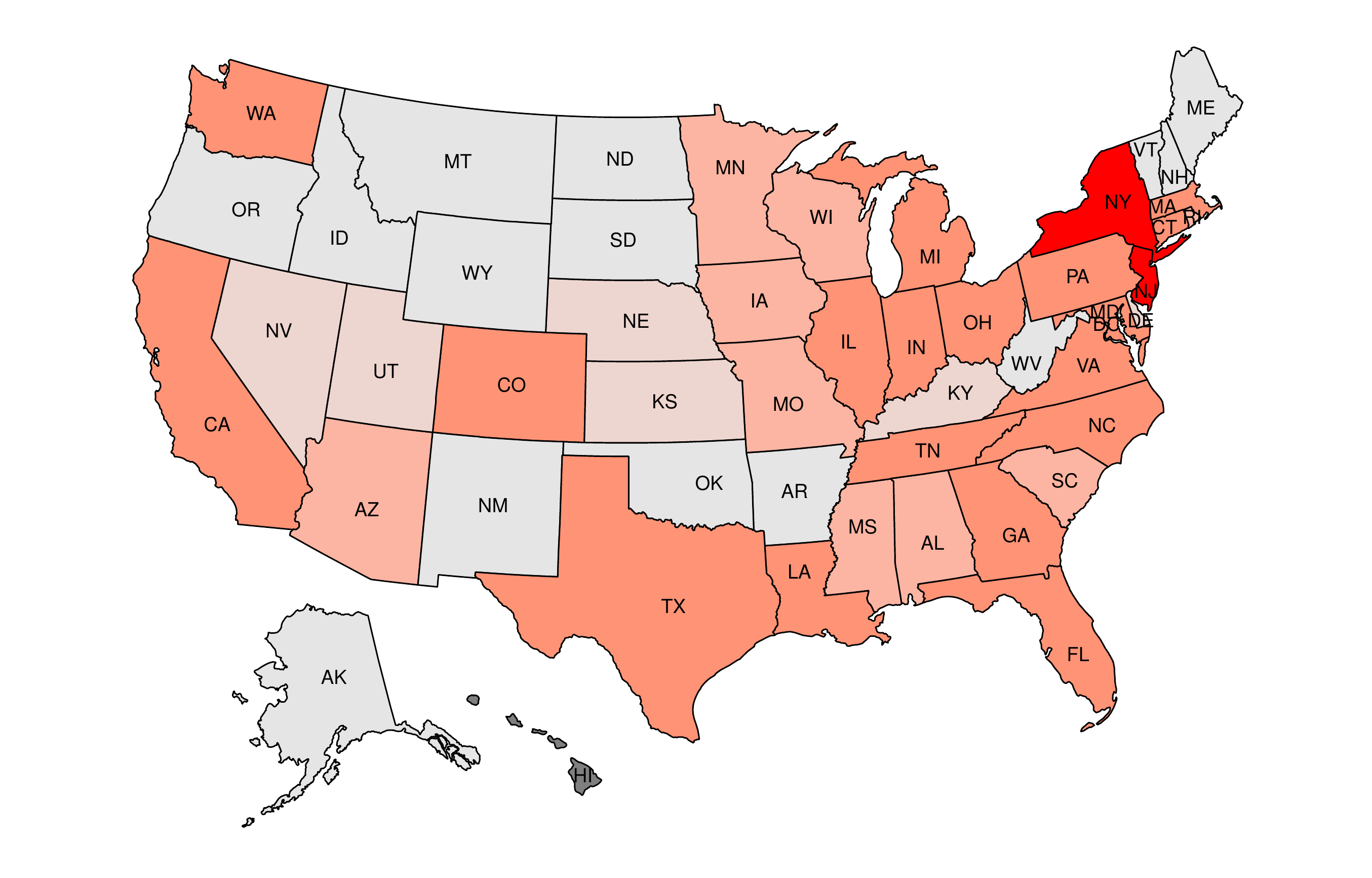}
	\caption{Clustering results based on the number of cumulative 
		confirmed cases in 50 continental states in the US from 
		04/01/2020 to 05/15/2020.}
	\label{Fig:clusterhalf}
\end{figure}

For comparison, we also conduct  an analogous analysis on a 
truncated study 
period from 04/01/2020 to 05/15/2020, around which business 
reopening 
starts over a large number of states. The clustering 
results are presented in Figure~\ref{Fig:clusterhalf}, and the 
cluster 
structure appears to be greatly different from that given in 
Figure~\ref{Fig:cluster}. Essentially, the New York state and its 
neighbors
are the epidemic centers for the period from 04/01/2020 to 
05/15/2020, and 
the alterations in the cluster membership flag the significant 
impacts of reopening
the business in accelerating the spread of the virus.

\section{Forecasting}
\label{sec:forcasting}

When forecasting the COVID-19 data using functional data approaches, 
we need to preserve the temporal dynamics among curves to maintain 
the forecasting ability, which leads us to the context of {\em 
	functional time series}~\cite[FTS,][]{Hormann2010weakly}. 

\subsection{Converting data into FTS}
\label{subsec:ftsconversion}
An FTS consists of a set of random functions collected over time. So 
far in the literature, there are two common formations of an FTS 
object. One treats an FTS as a segmentation of an almost continuous 
time record into natural consecutive intervals, such as days, 
months, or quarters. This is a conventional treatment for financial 
data~\cite{Hormann2012functional}.
The other treats it as a collection of curves over a time 
period, where the curves are not functions of time; 
see~\cite{Chiou2009modeling} for example. The major difference 
between 
these two formations is whether the 
continuum of each function is a time variable. In this section, we 
start with the number of nationwide cumulative confirmed cases of 
the COVID-19 in the US, 
where we define and convert the FTS using the first convention 
outlined above. 
Based on several reports on the COVID-19 
pandemic~\cite{Mcaloon2020, Zhang2020}, 
the average incubation period of the coronavirus is about $7$ days, 
but the incubation period could last up to $14$ days for most of the 
cases. Therefore, we segment the number of cumulative confirmed 
cases into curves with time intervals of $14$ days. In line with the 
conventional treatment that segments a continuous curve into small 
intervals, we set the last value in one curve equal to the first 
value of the next.

Similar to the classical time series modeling, it is critical to 
appropriately handle the non-stationary data when modeling the FTS. 
To circumvent the issue of the non-stationarity in the COVID-19 data,
we exploit the ideas from the literature dealing with FTS of share 
prices. In~\cite{Gabrys2010tests}, the authors define the cumulative 
intraday returns (CIDR's), and later in~\cite{Horvath2014testing}, a 
formal test that justifies the stationarity of CIDR's curves has been
developed. Here we convert the non-stationary cumulative confirmed 
case counts into stationary curves by calculating the daily growth 
rate in each curve:
\begin{equation}
	\label{eq:growthrate}
	r_{n,j} = 100 \times \bigl[\ln C_{n,j} - \ln C_{n,j-1} \bigr],
\end{equation}
where $C_{n,j}$ denotes the number of confirmed cases on the $j$-th 
day of the $n$-th segment, for $j \in \{1, 2, \ldots, 14\}$, and $n 
\in \{1, 2, \ldots, N\}$. 
Let $r_n{(t)}$ be the daily growth rates curve of the $n$-th 
segment, but values can only be observed at discrete time points 
$j$, such that $r_{n,j} = r_n{(t_j)}$.

Under the functional data framework, we assume the neighboring grid 
points are highly 
correlated, and smoothing serves as a tool for regularization, such 
that we borrow 
information from the neighboring grid 
points~\cite{Wang2016functional}. When applying to the daily growth 
rate 
for the cumulative confirmed cases, we assume that the 
underlying continuous and smooth function $Y_n(t)$ is observed at 
discrete points with smoothing error $\epsilon_n(t_j)$:
\begin{equation*}
	r_n(t_j) = Y_n(t_j) + \epsilon_n(t_j);
\end{equation*}
see~\cite{Hyndman2007robust} for details on smoothing. 
After applying the 
test from~\cite{Horvath2014testing}, we 
find that the $p$-value is equal to $0.989$, suggesting the 
stationarity of the FTS 
with respect to the daily growth rates.

The {\em rainbow plot} proposed in~\cite{Hyndman2010rainbow} is 
effective in
the visualization of the FTS. In a rainbow plot, 
functions that are ordered in time and colored with a spectrum 
of rainbow, such that functions from earlier times are colored in 
red, while the most recent ones are in violet. 
The rainbow plot captures the 
features of an FTS in two ways. Within each curve, the mode 
of variation reflects the pattern of the curve, and the ordering in 
color reveals the temporal dynamics over time.

\begin{figure}[!htbp]
	\centering
	\includegraphics[width=0.48\textwidth]{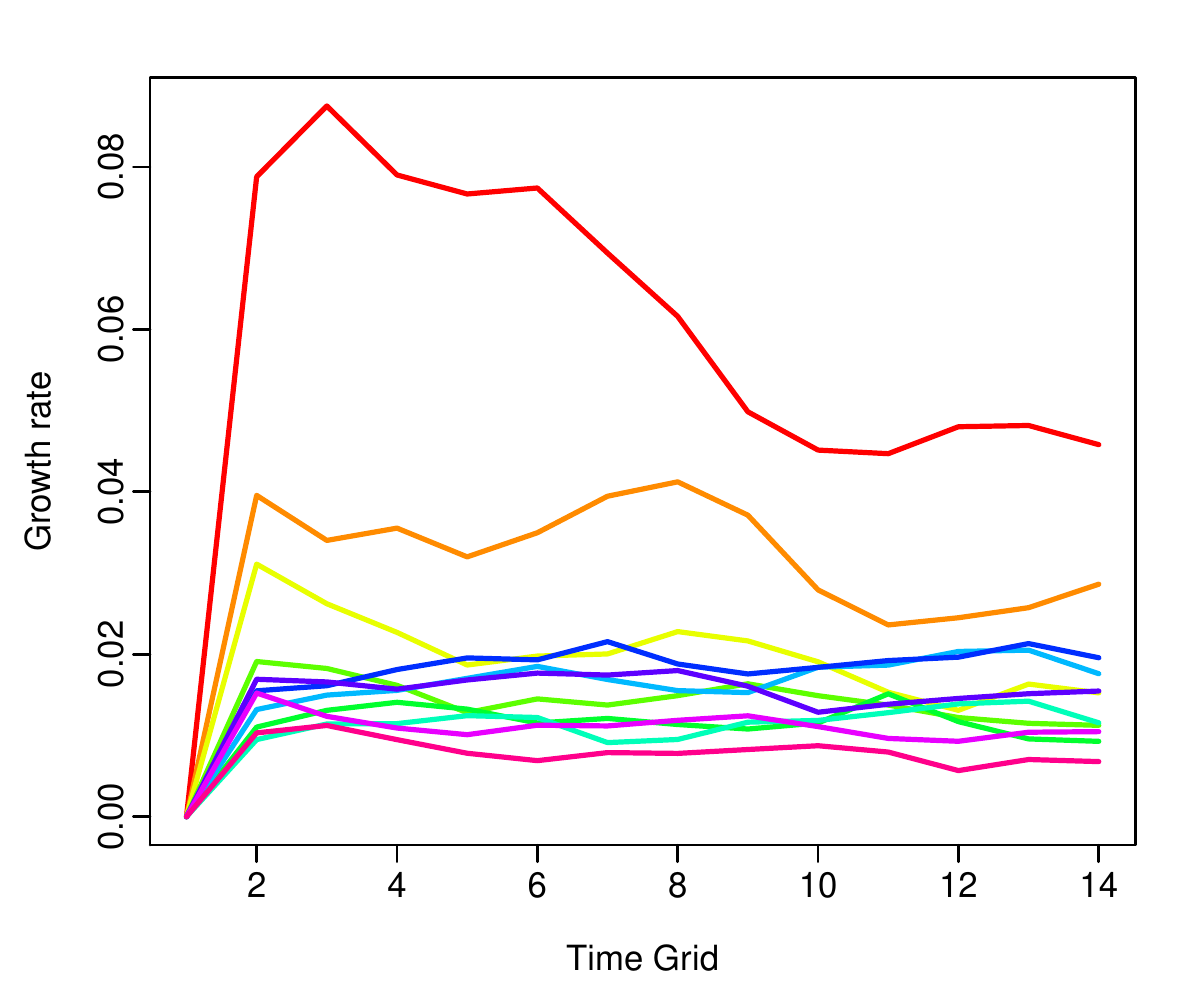}
	\includegraphics[width=0.48\textwidth]{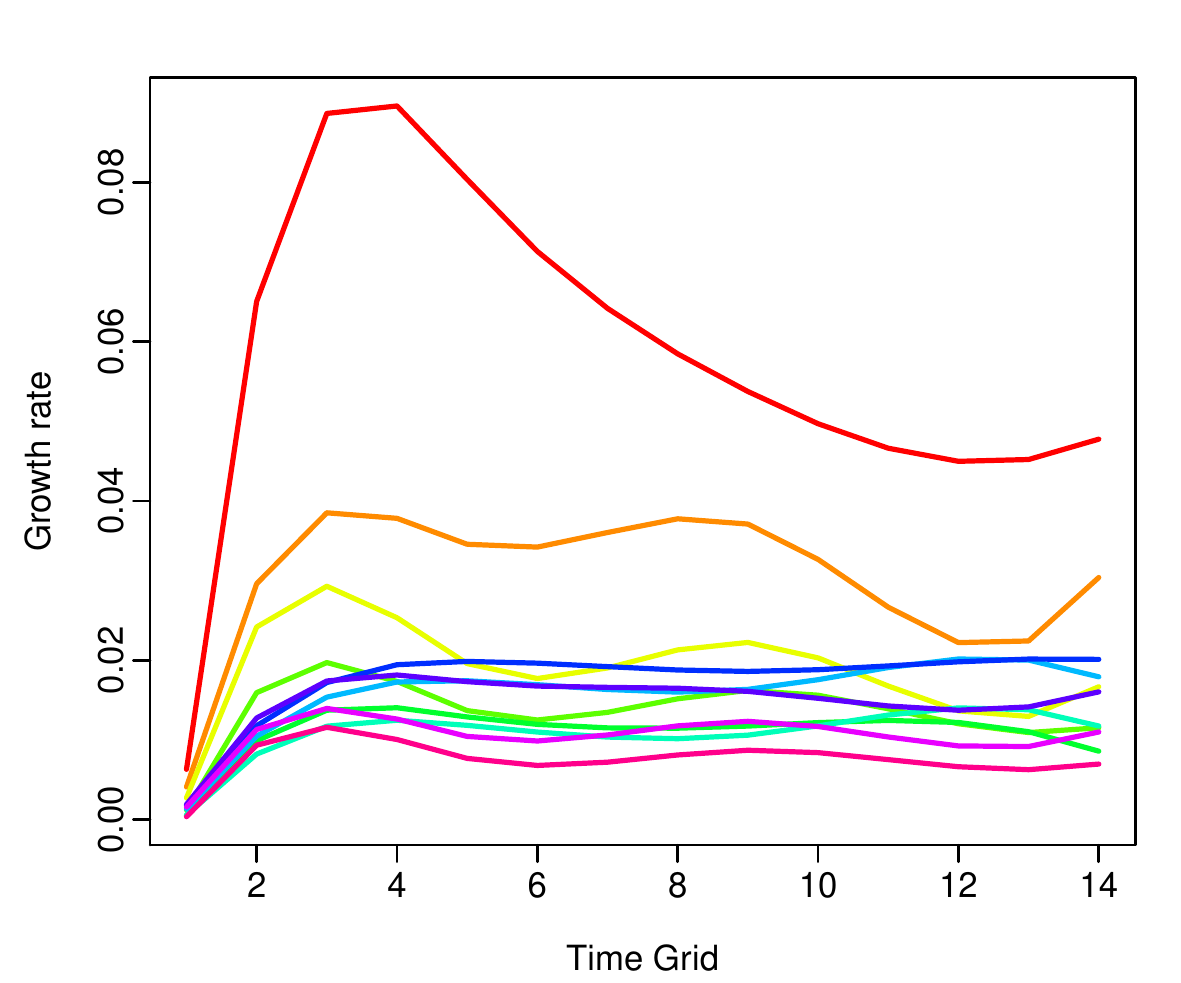}
	\caption{Functional time series for unsmoothed (left panel) 
		and smoothed (right panel) growth rates of the cumulative 
		confirmed cases of COVID-19 in the US.}
	\label{fig:rainbow}
\end{figure}

Figure~\ref{fig:rainbow} presents the rainbow plot of both 
unsmoothed 
(left) and smoothed (right) growth rates 
of the cumulative confirmed cases of COVID-19 in the US from 
04/04/2020 
to 
08/25/2020, leading to exactly $11$ functional curves. From 
Figure~\ref{fig:rainbow}, we observe that all 
the curves share a similar temporal pattern, where 
the growth rate increases rapidly within the first four days, and 
then tapers off gradually. This coincides with a scientific finding 
that an infected person is most contagious early in the course of 
their illness~\cite{Siordia2020}. Based on the color 
ordering in the rainbow plot, we see that the 
growth rate declines gradually from April to June, indicating the 
effectiveness of
the practice of public health measures (including lock down 
policies implemented at the early stage of the outbreak),  but it 
bounces back up again in July,
flagging the effects of reopening.

In the rest of this section, we develop a forecasting scheme by 
considering a dynamic FPCA method that accounts for the temporal 
dynamics of 
the long-run covariance structure of the FTS. The forecasting is 
done 
through the dynamic FPCA scores. 
We use the root mean squared forecasting error and a nonparametric 
bootstrap approach to assess the accuracy of 
point and interval 
forecasts, respectively. We then apply the proposed scheme to 
forecast the number of 
confirmed cases in the US in the next $13$ days (the first day of 
the $14$-day prediction segment coincides with the last day of the 
study period).

\subsection{Dynamic FPCA}
\label{subsec:dfpca}

The major critics of classical FPCA (c.f.\ 
Section~\ref{subsec:fpca}) stem from its incapability of making 
forecasts as it only aims to reduce the
dimension of data by maximizing the variances explained by 
eigenfunctions.
The dynamic FPCA, however, manages to reduce dimension towards the 
directions mostly reflecting temporal dynamics 
\cite{Hormann2015dynamic}. 
To better accommodate the need of 
capturing the temporal dynamics among the FTS and making better 
forecast, we here adopt the dynamic FPCA method 
\cite{Hormann2015dynamic}. 
The primary difference between classical 
and dynamic FPCA is that in performing the eigenanalysis, we 
replace the variance-covariance function with the long-run 
covariance 
function. 

For an FTS, the 
long-run covariance function, which is the functional analogue of 
the 
long-run covariance matrix in standard time series analysis, plays 
an important role in accommodating the temporal dependence among 
functions.  
We now give the definition of long-run 
covariance function.
Let $\{Y_i(\tau), i \in 
\mathbb{Z}\}$ denote a 
sequence of stationary and ergodic functional time series, 
with $\tau$ being a bounded continuous variable. The {\em 
	long-run covariance function} of $\{Y_i(\tau), i \in 
\mathbb{Z}\}$ is given by
\begin{equation*}
	\mathcal{C}(s, t) := \sum_{i = -\infty}^{\infty} \gamma_i(s, t),
\end{equation*}
where $\gamma_i(s, t) = \Cov(Y_0(s), Y_i(t))$. In practice, we need 
to
estimate $\mathcal{C}(s, t)$ via a finite sample, i.e. 
$\{Y_i(\tau), i = 1, 2, \ldots, n\}$ for some integer $n < \infty$. 
We use the {\em lag-window} estimator proposed 
by~\cite{Panaretos2013} to estimate
the long-run covariance function $\mathcal{C}(s, t)$:
\begin{equation}
	\label{eq:covkernel}
	\widehat{\mathcal{C}}(s, t) := \sum_{i = -\infty}^{\infty} 
	\mathcal{K} 
	\left(\frac{i}{h}\right) \widehat{\gamma}_{i}(s, t),
\end{equation}
where $\mathcal{K}(i/h)$ is a kernel function assigning 
different weights to the auto-covariance functions with different 
lags, and the parameter $h$ is the {\em 
	bandwidth}~\cite{andrews1991heteroskedasticity}. Typically, we 
assign more weights to the autocovariance functions of 
small lags and fewer to those of large lags. Here we use the 
``flat-top'' type of kernels, as they provide smaller bias and 
faster rates of convergence~\cite{politis1996flat}. For $k < 1$, 
the flat-top kernel is given by
\begin{equation*}
	\displaystyle
	K\left(\frac{i}{h}\right) = 
	\begin{cases}
		1, &\qquad 0 \le |i/h|< k;
		\\ \frac{|i / h|-1}{k-1}, &\qquad k \le |i/h| < 1;
		\\ 0, &\qquad |i/h| \ge 1.
	\end{cases}
\end{equation*}

The selection of the bandwidth crucially affects the accuracy of 
estimation. Here we adopt an adaptive 
bandwidth selection procedure~\cite{Rice2017plug} to get an 
approximately optimal bandwidth for the subsequent estimation of the 
long-run covariance of FTS. 
The 
estimator $\widehat{\gamma}_i(s, t)$ is given by
\begin{equation*}
	\widehat{\gamma}_i (s, t) = 
	\begin{cases}
		\frac{1}{n - i} \sum_{r = 1}^{n - i} Y_r(s) Y_{r + i}(t), 
		&\qquad i 
		\ge 0;
		\\ 
		\frac{1}{n - i} \sum_{r = 1 - i}^{n} Y_r(s) Y_{r + i}(t), 
		&\qquad i 
		< 0.
	\end{cases}
\end{equation*}
The dynamic FPCA is done through the Karhunen-Lo\`{e}ve expansion 
exactly the same as Equation \eqref{eq:fpca}, but the eigenvalues 
and eigenfunctions are estimated based on the 
eigenanalysis of $\widehat{\mathcal C}(s, t)$.

\subsection{Forecasting based on scores}
\label{subsec:forecasting}

We use $m$ functional observations to get an $\ell$-step-ahead 
forecast by the method developed 
in~\cite{Hyndman2009forecasting}. Note that 
applying a 
univariate time series forecasting 
method to the score vector $\bm{\widehat\xi}^{(m)}_k = 
\{\widehat\xi_{1k},\widehat\xi_{2k}
\ldots,\widehat\xi_{m k}\}$ gives the estimated $\widehat{\xi}_{m + 
	\ell \, | \, m, k}$, 
then the estimate of the $\ell$-step-ahead 
forecast is given by
\begin{equation*}
	\widehat Y_{m + \ell \,| \, m} (t) = \widehat\mu (t) + 
	\sum_{k = 1}^{K} \widehat\xi_{m + \ell \,| \, m,k} \,  
	\widehat\phi_k (t),
\end{equation*}
where $\widehat\mu(t)$ and $\{\widehat\phi_k(t), k = 1, 2, \ldots, 
K\}$ denote the estimated 
mean and FPCs, respectively.

For the application to the COVID-$19$ data, we forecast the daily 
number of the cumulative confirmed cases by the grid 
points on the forecasting curve. Without loss of generality, we 
consider the one-step-ahead forecast (i.e., $\ell = 1$). Note that 
this procedure generates 
daily forecasts up to $13$ days, since by our assumption on the FTS 
conversion in Section~\ref{subsec:ftsconversion}, the first grid 
point on the curve 
is the same as that on the last day of the study period.
By implementing the FTS converting 
procedures from~\ref{subsec:ftsconversion}, we obtain an FTS object 
consisting of 
$11$ functional curves. We create an {\em expanding window analysis} 
framework, and apply the proposed forecasting method to get multiple 
one-step-ahead forecasts on the growth rate curve. The details 
are deferred to Section~\ref{subsec:point}.

\subsection{Forecast accuracy evaluation}
\label{subsec:accuracy}

In this section, we introduce the methods that assess the 
prediction accuracy for point and interval 
forecasts. To make forecast for the number of confirmed cases in 
the US in the next $13$ days, we set $\ell = 1$ throughout the 
section.

\subsubsection{Point forecast}
\label{subsec:point}

For a point forecast, We use the {\em root mean square forecast 
	error} (RMSFE) to evaluate its accuracy:
\begin{equation}
	\label{eq:point_eval}
	{\rm RMSFE}(j) = 
	\sqrt{\frac{1}{N - n}\sum_{m=n}^{N - 1}\left[C_{m+1}(j) 
		- 
		\widehat C_{m+1}(j)\right]^2},
\end{equation}
where $N$ is the total number of segments in the FTS, $m$ is the 
number of curves used in forecasting, $C_{m+1}(j)$ 
is the number of confirmed cases at the ${j}$-th point on 
the ${(m+1)}$-th curve, and $\widehat 
C_{m+1}(j)$ is the corresponding forecast value. The RMSFE 
measures the discrepancy between the forecast and the actual 
value.

Given that there are $11$ functional curves in the study period,
we start with the 
first $n = 8$ curves to generate one-step-ahead point forecasts, and 
repeat this forecasting procedure by adding one curve at a time 
until the first $10$ curves are included, which gives us $(N - n = 
3)$ 
one-step-ahead forecasts in total. In other words, for each time 
point $j$, there are three forecast values. We choose $n = 8$ as our 
starting point to ensure that an adequate number of curves are used 
for forecasting.

\begin{figure}[ht]
	\centering
	\includegraphics[width=\textwidth]{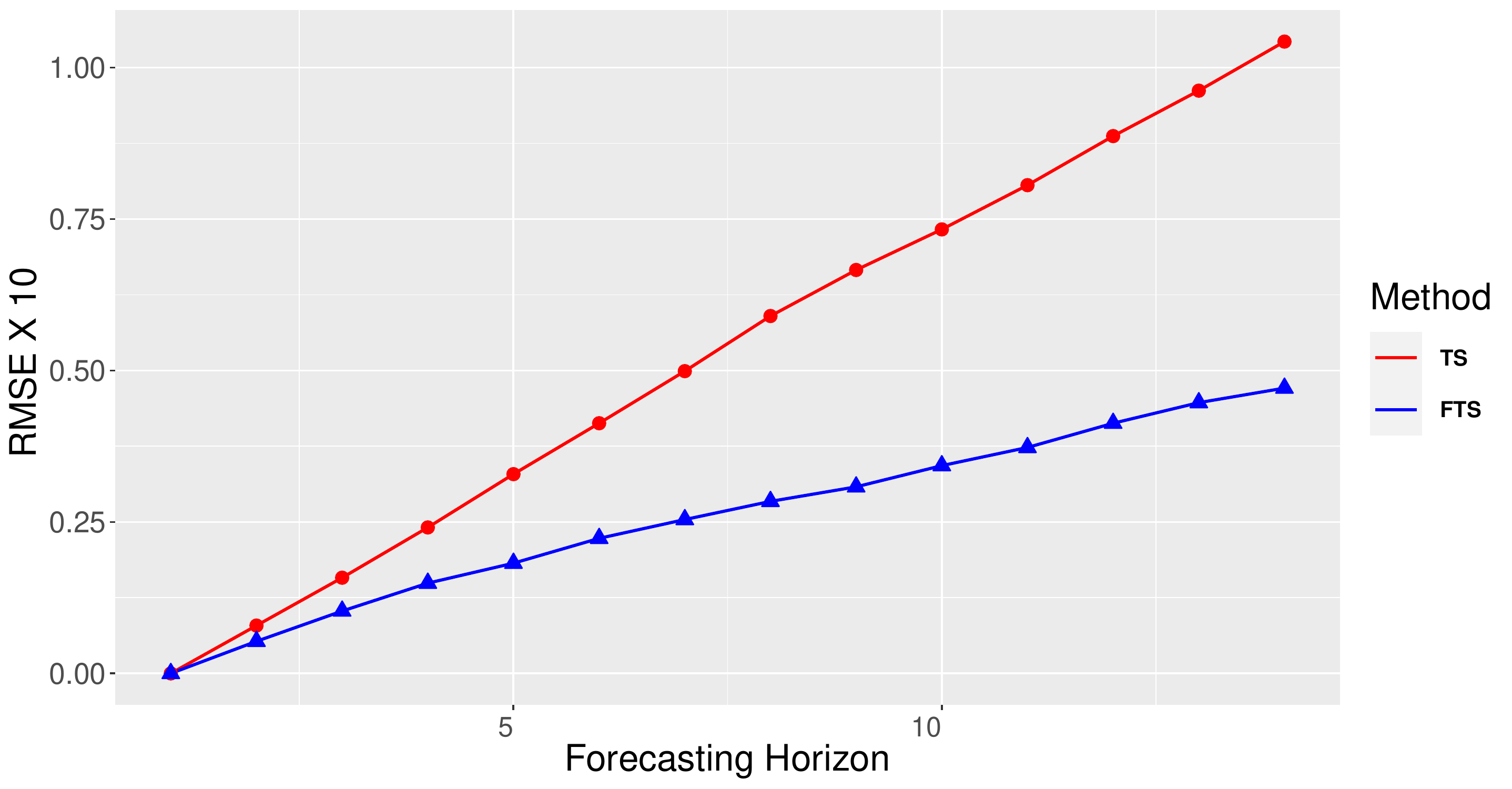}
	\caption{Average RMSFE values (scaled by $10$) for the FTS and 
		the standard ARIMA method for the next $13$ 
		days. The FTS method consistently outperforms the standard 
		ARIMA. Note that the RMSFE curve starts from 0, 
		corresponding to
		the assumption in Section~\ref{subsec:ftsconversion} that 
		the first day of 
		the $14$-day prediction segment coincides with the last day 
		of the 
		study period.}
	\label{fig:rmseplot}
\end{figure}

We adopt a standard forecasting approach through the autoregressive 
integrated moving average (ARIMA) model as a competing method. The 
forecasting results are presented in Figure~\ref{fig:rmseplot}, 
where the $y$-axis is the average of the RMSFE (scaled by $10$) of 
the three one-step-ahead point forecasts. We observe that the mean 
values 
of the RMSFE of FTS forecasts are consistently smaller than the 
counterparts of standard ARIMA, suggesting that the proposed FTS 
method is preferred. 

\subsubsection{Interval forecast}
\label{subsec:interval}

To better capture the uncertainty of point forecasts, we construct 
the associated prediction intervals in this section. 
In~\cite{Aue2015prediction}, uniform prediction intervals are 
generated
using a parametric method, and later \cite{Shang2018bootstrap} 
suggests
constructing pointwise prediction intervals via a 
nonparametric bootstrap method. 
The construction of the prediction 
intervals is based 
on in-sample-forecast errors, i.e., $\widehat 
e_{m +1 \,| \, m}(j) = C_{m + 1}(j)-\widehat 
C_{m + 1}(j)$. Specifically, we generate a bootstrap sample 
(with replacement) of forecasting errors to obtain the upper and 
lower bounds, respectively denoted by $\eta^{\rm ub}(j)$ and 
$\eta^{\rm up}(j)$. Then we choose a tuning 
parameter, $\delta_\alpha$, such that 
\begin{equation*}
	\Expp \left\{\delta_\alpha \times \eta^{\rm lb}(j)\leq 
	\widehat e_{m +1 \,| \, m} \leq \delta_\alpha \times \eta^{\rm 
		ub}(j)\right\} = (1-\alpha) \times 100\%.
\end{equation*}
The one-step-ahead (pointwise) prediction interval is given by
\begin{equation}
	\label{eq:PI}
	\widehat C_{m + 1}(j) + \delta_\alpha \times \eta^{\rm 
		lb}(j) \leq C_{m + 1}(j) \leq 
	\widehat C_{m + 1}(j) + \delta_\alpha \times 
	\eta^{\rm ub}(j).
\end{equation}

\begin{figure}[ht]
	\centering
	\includegraphics[width=\textwidth]{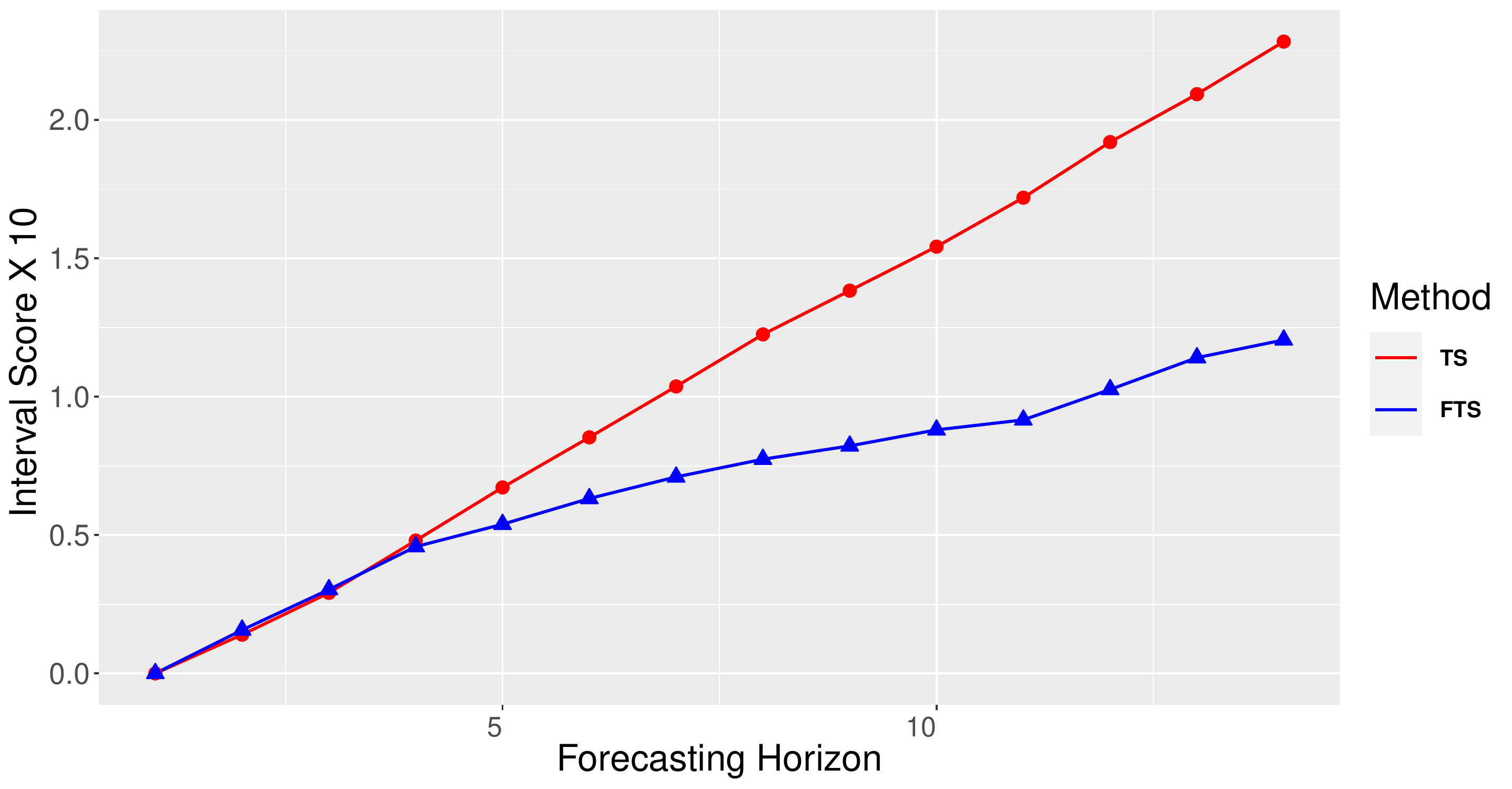}
	\caption{Average interval scores (scaled by $10$) for the 
		FTS method and the standard ARIMA model 
		for the next $13$ days. The FTS approach is preferred in 
		that it produces smaller 
		interval scores in most of the cases.
		Note that the RMSFE curve starts from 0, corresponding to
		the assumption in Section~\ref{subsec:ftsconversion} that 
		the first day of 
		the $14$-day prediction segment coincides with the last day 
		of the 
		study period.}
	\label{fig:scoreplot}
\end{figure}

We use the interval scoring rules~\cite{Gneiting2007strictly} to 
evaluate the accuracy of the pointwise interval forecasts. The 
interval score for the pointwise interval forecast (c.f.\ 
Equation~\eqref{eq:PI}) at time point 
$j$ is given by
\begin{equation*}
	\begin{split}
		&S_\alpha\left[\widehat C_{m + 1}^{\rm lb}(j), \widehat 
		C_{m + 1}^{\rm ub}(j); C_{m+1}(j)\right] 
		\\ &= 
		\left[\widehat C_{m + 1}^{\rm ub}(j) - \widehat C_{m + 
			1}^{\rm lb}(j)\right] + \frac{2}{\alpha}\left[\widehat 
		C_{m 
			+ 1}^{\rm 
			lb}(j) - 
		\widehat C_{m + 1}(j)\right]\bm{1}\left\{ \widehat C_{m 
			+ 1}(j) < \widehat C_{m + 1}^{\rm lb}(j)\right\}
		\\ 
		&\quad{} + \frac{2}{\alpha}\left[\widehat C_{m + 1}(j) - 
		\widehat 
		C_{m + 1}^{\rm ub}(j) \right]\bm{1}\left\{ \widehat 
		C_{m + 1}(j) > \widehat C_{m + 1}^{\rm 
			ub}(j)\right\},
	\end{split}
\end{equation*}
where 
$\bm{1}\left\{\cdot\right\}$ represents a standard indicator 
function, and the level of significance, $\alpha$, is conventionally 
set at~$0.2$. A lower interval score suggests that the interval 
forecast is more accurate. The ideal scenario is that the actual 
values of $C_{m + 1}(j)$ values lie 
between $\widehat 
C_{m+1}^{\rm lb}(j)$ and $\widehat 
C_{m+1}^{\rm ub}(j)$ for all $j$. The mean interval score over $(N - 
n)$ 
one-step-ahead forecasts becomes
\begin{equation*}
	\bar{S}_\alpha(j) = 
	\frac{1}{N-n}\sum_{m=n}^{N-1}S_\alpha\Big[\widehat 
	C_{m+1}^{\text{lb}}(j), \widehat C_{m+1}^{\text{ub}}(j); 
	C_{m+1}(j)\Big].
\end{equation*}
Figure~\ref{fig:scoreplot} shows the mean interval 
scores (scaled by $10$) of the interval forecasts obtained from the 
FTS model versus the standard ARIMA model. The computation results 
reveal that the proposed FTS method outperforms the standard ARIMA.

\subsubsection{Future forecasts}
\label{subsec:future}

\begin{figure}[ht]
	\centering
	\includegraphics[width=\textwidth]{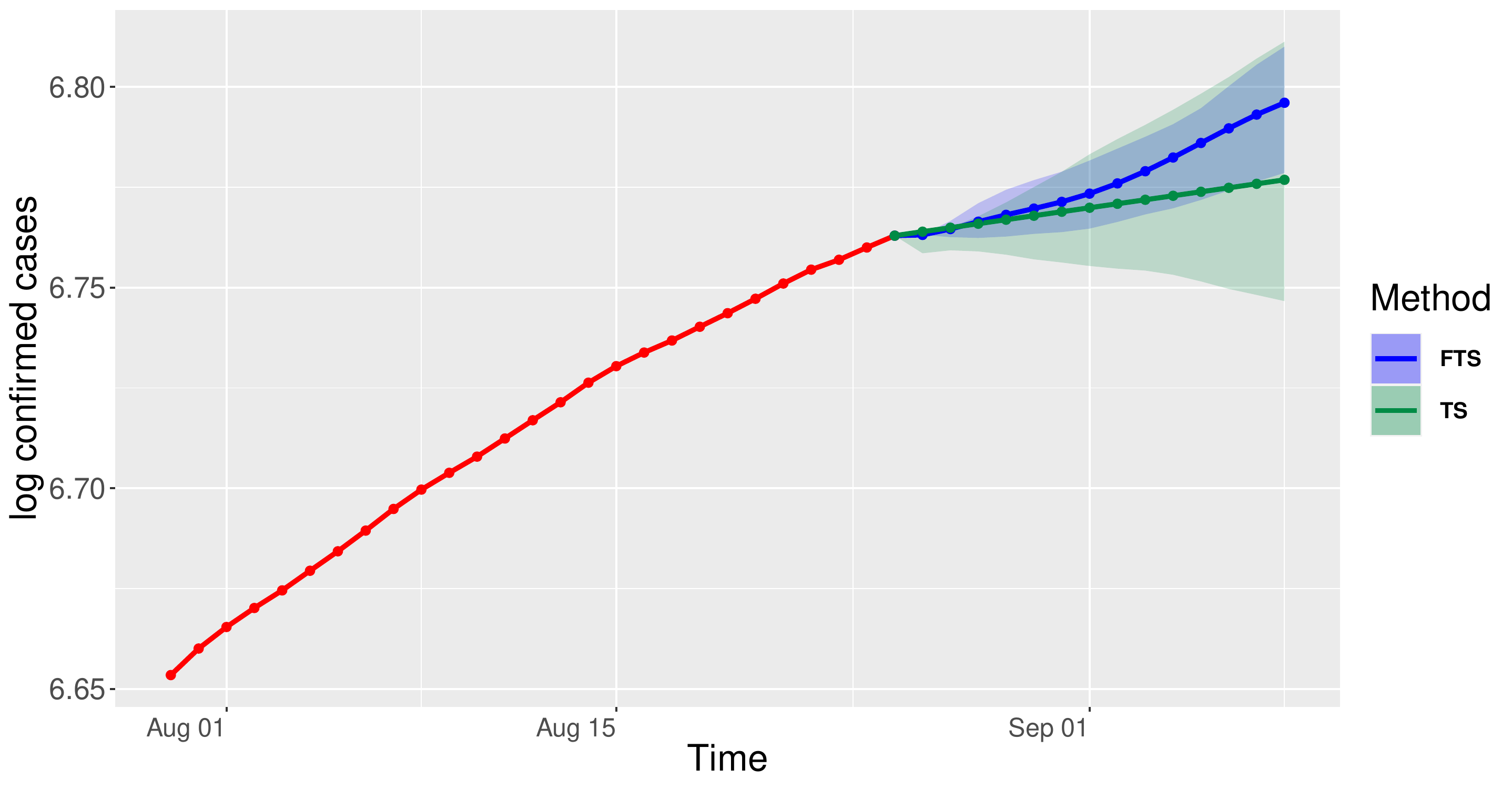}
	\caption{Point forecast and confidence bands for the number of 
		total 
		confirmed cases in the next 13 days (upon the last date of 
		the study 
		period)}
	\label{fig:future}
\end{figure}

We apply the proposed method to forecast the number of cumulative 
confirmed cases in the next $13$
days upon the 
last date of study period. Table~\ref{tab:future} shows the 
predicted 
number and 80\% confidence interval of nationwide cumulative 
confirmed cases (in thousands). In 
addition, the point forecasts and the associated 80\% confidence 
bands for the cumulative confirmed cases in the US from 08/26/2020 
to 
09/07/2020 are depicted in Figure~\ref{fig:future}, where we also 
compare the prediction results with those generated from an ARIMA 
model.

From Table~\ref{tab:future}, we see that our FTS forecasts tends to 
slightly underestimate the total counts, but the actual count still 
falls within the prediction intervals. Such deviation may be due to 
the reopening of the schools starting from mid-August, leading to 
the surge in the cumulative confirmed cases nationwide. When 
comparing the FTS forecasting results with the ARIMA model in 
Figure~\ref{fig:future}, we see a narrower prediction interval for 
the FTS results, suggesting that it is preferred than the standard 
ARIMA 
model.

\begin{table}[ht] \centering 
	\caption{Number of cumulative confirmed cases (in thousands) in 
		the United 
		States in the next 13 days (upon the last date of the study 
		period).}
	\label{tab:future} 
	\begin{tabular}{lccccccc}
		\hline
		Date & 08/26 & 08/27 & 08/28 & 08/29 & 08/30 & 08/31 & 
		09/01 \\
		\hline
		Actual & $5,839$ & $5,884$ & $5,931$ & $5,975$ & $6,009$ &  
		$6,045$ & $6,089$\\
		\hline 
		Forecasts  & $5,815$ & $5,840$ & $5,863$ & $5,884$ & 
		$5,907$ & $5,935$ & $5,970$\\
		Lower Bound  & $5,788$ & $5,786$ & $5,790$ & $5,799$ & 
		$5,805$ & $5,817$ &$5,839$ \\
		Upper Bound  & $5,843$ & $5,902$ & $5,948$ & $5,981$ & 
		$6,010$ &  $6,049$ & $6,090$\\
		\hline
		Date  & 09/02 & 09/03 & 09/04 & 09/05 & 09/06 & 09/07 &\\ 
		\hline
		Actual  & $6,122$ & $6,168$ & $6,220$ & $6,263$ & $6,293$ &  
		$6,318$&\\
		\hline
		Forecasts  & $6,012$ & $6,060$ & $6,110$ & $6,161$ & 
		$6,210$ & $6,252$ &\\
		Lower Bound  & $5,865$ & $5,885$ & $5,913$ & $5,946$ & 
		$5,976$ & $6,006$ & \\
		Upper Bound  & $6,131$ & $6,176$ & $6,233$ & $6,312$ & 
		$6,390$ &  $6,457$ & \\
		\hline
	\end{tabular} 
\end{table}

\section{Discussions}
\label{sec:discus}

In this article, we conduct a functional data analysis of the time 
series data of COVID-19 in the US. 
Based on our results, it is evident that the practice of public 
health measures (e.g. ``stay-at-home'' order and mask wearing) helps 
reduce the growth rate of the epidemic outbreak 
over the nation. However, the implementation of the business 
reopening plans seems to have caused the rapid spread of the 
COVID-19 in 
some states, e.g. Texas and Florida. 

We quantitatively assess the correlation between confirmed and death 
cases using the FCCA. Overall, we observe a high canonical 
correlation between confirmed and death cases, though the canonical 
variable scores vary from state to state. 
With the population size in each state carefully adjusted, we see 
that 
there is a substantial 
change in the cluster structure in the early and late times, and 
05/15/2020 (the average date of business reopening across the 
states) 
appears to be a potential change point. Besides, states that 
are geographically close to the hot spots are likely to be 
clustered together, and population density (at state level) appears 
to 
be a critical factor affecting the cluster structure.

In addition, we also propose a forecasting scheme for the nationwide 
cumulative 
confirmed cases under the functional time series framework. 
Integrating
information from the neighboring data point, the forecasting 
accuracy from the functional time series approach outperforms that 
from an ARIMA model. 
Forecasts are also 
made to the next $13$ days of the study period, and comparisons with 
the actual
counts are provided. Although our method tends to produce smaller 
estimated counts for the next 13 days,
the actual values still fall within the prediction intervals 
associated with our forecasts.
It is also worthwhile noticing that such underestimation may 
indicate the 
effects of school reopening in accelerating the spread of the virus 
over the country.

\section*{Acknowledgement}
The authors would 
like to thank Drs.\ Jun Yan from the University of Connecticut and 
Zifeng Zhao from the University of Notre Dame for valuable comments 
on the manuscript.


\begin{thebibliography}{9}
	
	\bibitem{Abdollahi2020}
	\textsc{Abdollahi, E., Champredon, D., Langley, J. M., Galvani, 
		A. P.} and \textsc{Moghadas, S. M.} (2020). Temporal 
		estimates 
	of case-fatality rate for COVID-19 outbreaks in Canada and the 
	United States. \textit{Can. Med. Assoc. J.} \textbf{192} 
	E666--E670.
	
	\bibitem{Abraham2003}
	\textsc{Abraham, C., Cornillon, P. A., Matzner-L{\o}ber, E.} and 
	\textsc{Molinari, N.} (2003). Unsupervised curve clustering 
	using 
	B-splines. \textit{Scand. J. Statist.} \textbf{30} 581--595. 
	\MR{2002229}
	
	\bibitem{andrews1991heteroskedasticity}
	\textsc{Andrews, D.} (1991). Heteroskedasticity and 
	autocorrelation consistent covariance matrix estimation. 
	\textit{Econometrica} \textbf{59} 817--858.MR{1106513} 
	
	\bibitem{Aue2015prediction}
	\textsc{Aue, A., Norinho, D. D.} and 
	\textsc{H\"{o}rmann, S.} (2015). On the prediction of stationary 
	functional time series. \textit{J. Amer. Statist. Assoc.} 
	\textbf{110} 378--392. 
	\MR{3338510}
	
	\bibitem{Biernacki2003}
	\textsc{Biernacki, C., Celeux, G.,} and \textsc{Govaert, G.}. 
	(2003). Choosing starting values for the EM algorithm for 
	getting the highest likelihood in multivariate Gaussian mixture 
	models. \textit{Comput. Statist. Data Anal.} \textbf{41} 
	561--575. \MR{1968069}
	
	\bibitem{Capra1991}
	\textsc{Capra, W. B.} and \textsc{M\"{u}ller, H.-G.} (1991). An 
	accelerated-time model for response curves. \textit{J. Amer. 
		Statist. Assoc.} \textbf{92} 72--83. \MR{1436099}
	
	\bibitem{Carroll2020}
	\textsc{Carroll, C., Gajardo, A., Chen, Y. Dai, X., Fan, J. 
		Hadjipantelis, P. Z., Han, K., Ji, H., M\"{u}eller, H.-G.} 
		and 
	\textsc{Wang, J.-L.} (2020). fdapace: Functional Data Analysis 
	and Empirical Dynamics. R package version 0.5.3, 
	\url{https://CRAN.R-project.org/package=fdapace}.
	
	\bibitem{Castro1986}
	\textsc{Castro, P. E., Lawton, W. H.} and \textsc{Sylvestre, E. 
		A.} (1986). Principal modes of variation for processes with 
	continuous sample curves. \textit{Technometrics} \textbf{28} 
	329--337. 
	
	\bibitem{Chen2015}
	\textsc{Chen, W.-C.} and \textsc{Maitra, R.} (2015). EMCluster: 
	EM algorithm for 
	model-based clustering of finite mixture Gaussian distribution. 
	R Package, \url{http://cran.r-project.org/package=EMCluster}.
	
	\bibitem{Chiou2007}
	\textsc{Chiou, J.-M.} and \textsc{Li, P.-L.} (2007). Functional 
	clustering and identifying substructures of longitudinal data. 
	\textit{J. R. Stat. Soc. Ser. B Stat. Methodol.} \textbf{69} 
	679--699. \MR{2370075}
	
	\bibitem{Chiou2008}
	\textsc{Chiou, J.-M.} and \textsc{Li, P.-L.} (2008). 
	Correlation-based functional clustering via subspace projection. 
	\textit{J. Amer. Statist. Assoc.} \textbf{103} 1684--1692. 
	\MR{2724685}
	
	\bibitem{Chiou2009modeling}
	\textsc{Chiou, J.-M.} and \textsc{M\"{u}ller, H.-G.} (2009). 
	Modeling hazard rates as functional data for the analysis of 
	cohort lifetables and mortality forecasting. \textit{J. Amer. 
		Statist. Assoc.} \textbf{104} 572--585. 
	\MR{2751439}
	
	\bibitem{Chung1997}
	\textsc{Chung, F. R. K.} (1997). \textit{Spectral Graph Theory}. 
	American Mathematical Society, Providence, RI. \MR{1421568}
	
	\bibitem{Dauxois1982}
	\textsc{Dauxois, J., Pousse, A.} and \textsc{Romain, Y.} (1982). 
	Asymptotic theory for the principal component analysis of a 
	vector random function: some applications to statistical 
	inference. \textit{J. Multivariate Anal.} \textbf{12} 136--154. 
	\MR{0650934}
	
	\bibitem{Dempster1977}
	{\textsc Dempster, A. P., Laird, N. M.} and \textsc{Rubin, D. 
		B.} (1977). Maximum likelihood from incomplete data via the 
		EM 
	algorithm. \textit{J. Roy. Statist. Soc. Ser. B} \textbf{39} 
	1--38. \MR{0501537}
	
	\bibitem{Fan1996}
	\textsc{Fan, J.} and \textsc{Gijbels, I.} (1996). \textit{Local 
		Polynomial Modelling and its Application}. Chapman \& Hall, 
	London, UK. \MR{1383587}
	
	\bibitem{Ferreira2009}
	\textsc{Ferreira, L.} and \textsc{Hitchcock, D. B.} (2009). A 
	comparison of hierarchical methods for clustering functional 
	data. \textit{Comm. Statist. Simulation Comput.} \textbf{38} 
	1925--1949. \MR{2751180}
	
	\bibitem{Gabrys2010tests}
	\textsc{Gabrys, R., Horv{\'a}th, L.} and \textsc{Kokoszka, P.} 
	(2010). Tests for error correlation in the functional linear 
	model. \textit{J. Amer. Statist. Assoc.} \textbf{105} 
	1113--1125. \MR{2752607}
	
	\bibitem{Gneiting2007strictly}
	\textsc{Gneiting, T.} and \textsc{Raftery, A. E.} (2007). 
	Strictly proper scoring rules, prediction, and estimation. 
	\textit{J. Amer. Statist. Assoc.} \textbf{102} 359--378. 
	\MR{2345548}
	
	\bibitem{Grubaugh2020}
	\textsc{Grubaugh, N. D., Hanage, W. P.} and \textsc{Rasmussen, 
		A. L.} (2020). Making sense of mutation: What D614G means 
		for 
	the COVID-19 pandemic remains unclear. \textit{Cell} 
	\url{https://doi.org/10.1016/j.cell.2020.06.040}.
	
	\bibitem{hansen1982large}
	\textsc{Hansen, L. P.}
	(1982). Large sample properties of generalized method of moments 
	estimators. 
	\textit{Econometrica} \textbf{50}, 1029--1054. \MR{0666123}
	
	
	\bibitem{He2003}
	\textsc{He, G., M\"{u}ller, H.-G.} and \textsc{Wang, J.-L.} 
	(2003). Functional canonical analysis for square integrable 
	stochastic processes. \textit{J. Multivariate Anal.} \textbf{85} 
	54--77. \MR{1978177}
	
	\bibitem{He2004}
	\textsc{He, G., M\"{u}ller, H.-G.} and \textsc{Wang, J.-L.} 
	(2004). Methods of canonical analysis for functional data. 
	\textit{J. Statist. Plann. Inference} \textbf{122}, 141--159. 
	\MR{2057919}
	
	\bibitem{He2020}
	\textsc{He, W., Grace, Y. Y.} and \textsc{Zhu, Y.} (2020). 
	Estimation of the basic reproduction number, average incubation 
	time, asymptomatic infection rate, and case fatality rate for 
	COVID‐19: Meta‐analysis and sensitivity analysis. \textit{J. 
		Med. Virol.} \url{https://doi.org/10.1002/jmv.26041}.
	
	\bibitem{Heetal2020}
	\textsc{He, X., Lau, E.H.Y., Wu, P. et al.} (2020). Temporal 
	dynamics in viral shedding and transmissibility of COVID-19. 
	\textit{Nat. Med.}. \textbf{26}, 672--675. 
	
	\bibitem{Hormann2010weakly}
	\textsc{H\"{o}rmann, S.} and \textsc{Kokoszka, P.} 
	(2010). Weakly dependent functional data. 
	\textit{Ann. Statist.} \textbf{38}, 1845--1884. 
	\MR{2662361}
	
	\bibitem{Hormann2012functional}
	\textsc{H\"{o}rmann, S.} and \textsc{Kokoszka, P.} 
	(2010). Functional time series, in T.~S. Rao, S.~S. Rao
	and C.~Rao (Eds.) \textit{Handbook of Statistics} 
	\textbf{30} 157--186, Elsevier, Amsterdam, North Holland.
	
	\bibitem{Hormann2015dynamic}
	\textsc{H\"{o}rmann, S., Kidzi\'{n}ski, {\L}.} and 
	\textsc{Hallin, M.} 
	(2015). Dynamic functional principal components. 
	\textit{J. R. Stat. Soc. Ser. B. Stat. Methodol.} \textbf{77}, 
	319--348. \MR{3310529}
	
	\bibitem{Horvath2012estimation}
	\textsc{Horv{\'a}th, L., Kokoszka, P.}, and \textsc{Reeder, R.} 
	(2012). Estimation of the mean of functional time series and a 
	two-sample problem. \textit{J. R. Stat. Soc. Ser. B. Stat. 
		Methodol.} \textbf{75} 103--122. \MR{3008273}
	
	\bibitem{Horvath2014testing}
	\textsc{Horv{\'a}th, L., Kokoszka, P.}, and \textsc{Rice, G} 
	(2014). Testing stationarity of functional time series. 
	\textit{J. Econometrics} \textbf{179} 66--82. \MR{3153649}
	
	\bibitem{Hyndman2007robust}
	\textsc{Hyndman, R. J.}, and \textsc{Ullah, M. S.} 
	(2007). Robust forecasting of mortality and fertility rates: {A} 
	functional data approach. 
	\textit{Comput. Statist. Data Anal.} \textbf{51} 4942--4956. 
	\MR{2364551}
	
	\bibitem{Hyndman2009forecasting}
	\textsc{Hyndman, R. J.}, and \textsc{Shang, H. L.} 
	(2009). Forecasting functional time series. 
	\textit{J. Korean Statist. Soc.} \textbf{38} 219--221. 
	\MR{2750317}
	
	\bibitem{Hyndman2010rainbow}
	\textsc{Hyndman, R. J.}, and \textsc{Shang, H. L.} 
	(2010). Rainbow plots, bagplots, and boxplots for functional 
	data. 
	\textit{J. Comput. Graph. Statist.} \textbf{19} 29--45. 
	\MR{2752026}
	
	\bibitem{Hyndman2013coherent}
	\textsc{Hyndman, R.J., Booth, H.}, and \textsc{Farah, Y.} 
	(2013). Coherent mortality forecasting: the product-ratio method 
	with functional time series models. 
	\textit{Demography} \textbf{50} 261--283.
	
	\bibitem{Jacques2013}
	\textsc{Jacques, J.} and \textsc{Preda, C.} (2013). Funclust: A 
	curves clustering method using functional random variables
	density approximation. \textit{Neurocomputing} \textbf{112} 
	161--171.
	
	\bibitem{Jacques2014}
	\textsc{Jacques, J.} and \textsc{Preda, C.} (2014). Functional 
	data clustering: a survey. \textit{Adv. Data Anal. Classif.} 
	\textbf{8} 231--255. \MR{3253859}
	
	\bibitem{Jolliffe2002}
	\textsc{Jolliffe, I. T.} (2002). \textit{Principal Component 
		Analysis}, 2nd ed. Springer-Verlag, New York, NY.
	\MR{2036084}
	
	\bibitem{jhu:data}
	\textsc{Dong, E.}, \textsc{Du, H.}, and \textsc{Gardner, L.} 
	(2020). An interactive web-based dashboard to track COVID-19 in 
	real time. \textit{The Lancet infectious diseases}, 
	\textbf{20}(5), 533--534.
	
	\bibitem{Jones1992}
	\textsc{Jones, M. C.} and \textsc{Rice, J. A.} (1992). 
	Displaying 
	the important features of large collections of similar curves. 
	\textit{Amer. Statist.} \textbf{46} 140--145.
	
	\bibitem{Karhunen1946spektraltheorie}
	\textsc{Karhunen, K.}. (1946). Zur spektraltheorie 
	stochastischer prozesse. \textit{Ann. Acad. Sci. Fennicae Ser. 
		A. I. Math.-Phys.} 34. \MR{0023012}
	
	\bibitem{Kaufman1990}
	\textsc{Kaufman, L.} and \textsc{Rousseeuw, P. J.} (1990). 
	\textit{Finding Groups in Data: An Introduction to Cluster 
		Analysis}, 1st ed. Wiley-Interscience, Hoboken, NJ.
	\MR{2036084}
	
	\bibitem{Kosambi1943}
	\textsc{Kosambi, D. D.} (1943). Statistics in function space. 
	\textit{J. Indian Math. Soc.} \textbf{7} 76--88. \MR{0009816}
	
	\bibitem{Lee2012}
	\textsc{Lee, G.} and \textsc{Scott, C.} (2012). EM algorithms 
	for multivariate Gaussian mixture models with truncated and 
	censored data. \textit{Comput. Statist. Data Anal.} \textbf{56} 
	2816--2829. \MR{2915165}
	
	\bibitem{Leurgans1993}
	\textsc{Leurgans, S. E., Moyeed, R. A.} and \textsc{Silverman, 
		B. W.} (1993). Canonical correlation analysis when the data 
		are 
	curves. \textit{J. Roy. Statist. Soc. Ser. B} \textbf{55} 
	725--740. \MR{1223939} 
	
	\bibitem{Li2010}
	\textsc{Li, Y.} and \textsc{Hsing, T.} (2010). 
	Uniform convergence rates for nonparametric regression and 
	principal component analysis in functional/longitudinal data. 
	\textit{Ann. Statist.} \textbf{38} 3321--3351. 
	\MR{2541589}
	
	\bibitem{Li2013selecting}
	\textsc{Li, Y., Wang, N.}, and \textsc{Carroll, R. J.}. (2013). 
	Selecting the number of principal components in functional data. 
	\textit{J. Amer. Statist. Assoc.} \textbf{108} 1284--1294. 
	\MR{3174708}
	
	\bibitem{Liu2009}
	\textsc{Liu, B.} and \textsc{M\"{u}ller, H.-G.} (2009). 
	Estimating derivatives for samples of sparsely observed 
	functions, with application to online auction dynamics. 
	\textit{J. Amer. Statist. Assoc.} \textbf{104} 704--717. 
	\MR{2541589}
	
	\bibitem{Loeve1955probability}
	\textsc{Lo{\`e}ve, M.}. (1955). \textit{Probability Theory: 
		Foundations, Random Sequences}. D. Van Nostrand, Company, 
		Inc., 
	Princeton, NJ.
	
	
	\bibitem{Mcaloon2020}
	\textsc{McAloon, C., Collins, \'{A}, Hunt, K.} et al. (2020). 
	Incubation period of COVID-19: A rapid
	systematic review and meta-analysis of observational
	research. \textit{BMJ Open}, \textbf{10} e039652.
	
	\bibitem{McLachlan2019}
	\textsc{McLachlan, G.} and \textsc{Peel, D.} (2000). 
	\textit{Finite Mixture Models}. Wiley-Interscience, New York. 
	\MR{1789474}
	
	\bibitem{newey1987simple}
	\textsc{Newey, W. K.} and \textsc{West, K. D.} 
	(1987). A simple, positive semi-definite, heteroskedasticity and 
	autocorrelation consistent covariance matrix. 
	\textit{Econometrica} \textbf{55} 703--708. \MR{0890864}
	
	
	\bibitem{Omer2020}
	\textsc{Ommer, S. B., Malani, P.} and \textsc{del Rio, C.} 
	(2020). The COVID-19 andemic in the US: A clinical update. 
	\textit{JAMA} \textbf{323} 1767--1768.
	
	
	\bibitem{Panaretos2013}
	\textsc{Panaretos, V. M.} and \textsc{Tavakoli, S.} 
	(2013). Fourier analysis of stationary time series in function 
	space. \textit{Ann. Statist.} \textbf{41} 568--603. \MR{3099114}
	
	\bibitem{Peirlinck2020}
	\textsc{Peirlinck, M., Linka, K., Costabal, F. S.} and 
	\textsc{Kuhl, E.} (2020). Outbreak dynamics of COVID-19 in China 
	and the United States. \textit{Biomech. Model. Mechanobiol.} 
	\url{https://doi.org/10.1007/s10237-020-01332-5}.
	
	\bibitem{Peng2008}
	\textsc{Peng, J.} and \textsc{M\"{u}ller, H.-G.} (2008). 
	Distance-based clustering of sparsely observed stochastic 
	processes with applications to online auctions. \textit{Ann. 
		Appl. Stat.} \textbf{2} 1056--1077. \MR{2516804}
	
	\bibitem{politis1996flat}
	\textsc{Politis, D. N.} and \textsc{Romano, J. P.} (1996). 
	On flat-top kernel spectral density estimators for homogeneous 
	random fields. \textit{J. Statist. Plann. Inference} \textbf{51} 
	41--53. 
	
	\bibitem{Ramsay1982}
	\textsc{Ramsay, J. O.} (1982). When the data are functions. 
	\textit{Psychometrika} \textbf{47} 379--396. \MR{0691828}
	
	\bibitem{Ramsay1991}
	\textsc{Ramsay, J. O.} and \textsc{Dalzell, C. J.} (1991). Some 
	tools for functional data analysis. \textit{J. Roy. Statist. 
	Soc. 
		Ser. B} \textbf{53} 539--572. \MR{1125714}
	
	\bibitem{Ramsay2002}
	\textsc{Ramsay, J. O.} and \textsc{Silverman, B. W.} (2002). 
	\textit{Applied Functional Data Analysis: Methods and Case 
		Studies}. Springer-Verlag, New York, NY.
	\MR{1910407}
	
	\bibitem{Ramsay2020}
	\textsc{Ramsay, J. O., Graves, S.} and \textsc{Hooker, G.} 
	(2020). fda: Functional Data Analysis. R package version 5.1.4, 
	\url{https://CRAN.R-project.org/package=fda}.
	
	\bibitem{Rice2017plug}
	\textsc{Rice, G.} and \textsc{Shang, H. L.} (2017).  A plug-in 
	bandwidth selection procedure for long-run covariance estimation 
	with stationary functional time series. \textit{J. Time Series 
		Anal.} \textbf{38} 591--609. \MR{3664648}
	
	\bibitem{Shibata1981}
	\textsc{Shibata, R.} (1981) An optimal selection of regression 
	variables. \textit{Biometrika} \textbf{68} 45--54. \MR{0614940}
	
	\bibitem{Scotto2015}
	\textsc{Scotto, M. G., Wei$\beta$, C. H.} and \textsc{Gouveia, 
		S.} (2015). Thinning-based models in the analysis of 
	integer-valued time series: a review. \textit{Stat. Model.} 
	\textbf{15} 590--618. \MR{3441230}
	
	\bibitem{Shang2014}
	\textsc{Shang, H. L.} (2014). A survey of functional principal 
	component analysis. \textit{AStA Adv. Stat. Anal.} \textbf{98} 
	121--142. \MR{3254025}
	
	\bibitem{Shang2017forecasting}
	\textsc{Shang, H. L.}(2017). Forecasting intraday S\&P 500 index 
	returns: A functional time series approach, {\em Journal of 
	forecasting} \textbf{36} 741--755.
	
	\bibitem{Shang2018bootstrap}
	\textsc{Shang, H. L.} (2018). Bootstrap methods for stationary 
	functional time series. \textit{Stat. Comput.} \textbf{28} 
	1--10. \MR{3741633}
	
	\bibitem{Shang2019intraday}
	\textsc{Shang, H. L., Yang, Y.} and \textsc{Kearney, F.} 
	(2019).Intraday forecasts of a volatility index: functional time 
	series methods with dynamic updating. \textit{Ann. Oper. Res.} 
	\textbf{282} 331--354. \MR{4019243}
	
	\bibitem{Siordia2020}
	\textsc{Siordia, J. A. Jr.} (2020). Epidemiology and clinical 
	features of COVID-19: A review of current literature. \textit{J. 
		Clin. Virol.}, \textbf{127} 104357.
	
	\bibitem{Thorndike1953}
	\textsc{Thorndike, R. L.} (1953). Who belongs in the family? 
	\textit{Psychometrika} \textbf{18} 267--276.
	
	\bibitem{Ullah2013}
	\textsc{Ullah, S.} and \textsc{Finch, C. F.} (2013). 
	Applications 
	of functional data analysis: a systematic review. \textit{BMC 
		Med. Res. Methodol.} \textbf{13} 43. 
	
	\bibitem{Wang2016functional}
	\textsc{Wang, J.-L., Chiou, J.-M.} and \textsc{M\"{u}ller, 
	H.-G.} (2016). 
	Functional data analysis. \textit{Ann. Rev. Stat. Appl.} 
	\textbf{3} 257--295. 
	
	\bibitem{Yang2011}
	\textsc{Yang, W., M\"{u}ller, H.-G.} and 
	\textsc{Stadtm\"{u}ller, 
		U.} (2011). Functional singular component analysis. 
		\textit{J. R. 
		Stat. Soc. Ser. B Stat. Methodol.} \textbf{73} 303--324. 
	\MR{2815778} 
	
	
	\bibitem{Yao2005}
	\textsc{Yao, F., M\"{u}ller, H.-G.} and \textsc{ Wang, J.-L.} 
	(2005). Functional data analysis for sparse longitudinal data. 
	\textit{J. Amer. Statist. Assoc.} \textbf{100} 577--590. 
	\MR{2160561} 
	
	\bibitem{Yao2010}
	\textsc{Yao, F.} and \textsc{M\"{u}ller, H.-G.}	(2005). 
	Functional quadratic regression. 
	\textit{Biometrika} \textbf{97} 49--64. 
	\MR{2594416} 
	
	\bibitem{Zhang2016}
	\textsc{Zhang, X.} and \textsc{Wang, J.-L.} (2016). From sparse 
	to dense functional data and beyond. {\em Ann. Statist.} 
	\textbf{44} 2281--2321. \MR{3546451}
	
	\bibitem{Zhang2020}
	\textsc{Zhang, P., Wang, T.} and \textsc{Xie, S. X.} (2020). 
	Meta-analysis of several epidemic characteristics of COVID-19. 
	{\em J. Data Sci.} 
	\textbf{18} 536--549.

\end{thebibliography}
\end{document}